\documentclass[a4paper,fleqn]{cas-sc}

\usepackage[authoryear]{natbib}

\def\tsc#1{\csdef{#1}{\textsc{\lowercase{#1}}\xspace}}
\tsc{WGM}
\tsc{QE}

\usepackage[nointegrals]{wasysym}
\usepackage{amsmath}
\usepackage{amssymb} 
\usepackage{graphics}
\usepackage{dcolumn}

\newcommand{\colorindex}[2]{\filter{#1}$\!-\!$\filter{#2}}
\newcommand{\degsym}{^{\circ}}
\newcommand{\filter}[1]{$\mathit{#1}$}
\newcommand{\noteindex}[1]{\(^{\rm #1}\)}
\newcommand{\polenum}[1]{$P_{_#1}$}

\begin{document}
\let\WriteBookmarks\relax
\def\floatpagepagefraction{1}
\def\textpagefraction{.001}

\shorttitle{Spin vectors in the Koronis family: IV.}

\shortauthors{Slivan et al.}  

\title[mode = title]{Spin vectors in the Koronis family: IV.
Completing the sample of its largest members
after 35 years of study}

\author[adrMIT12,adrWellesley]{Stephen M. Slivan}[orcid=0000-0003-3291-8708]
\cormark[1]
\cortext[cor1]{Corresponding author}
\ead{slivan@mit.edu}

\author[adrWilliams]{Matthew Hosek Jr.}
\fnmark[1]
\fntext[1]{Present address Univ. of California at Los Angeles, CA, USA}

\author[adrColgate]{Max Kurzner}
\fnmark[2]
\fntext[2]{Present address Dept. of Physics and Astronomy, Univ. of Victoria, BC, Canada}

\author[adrColgate]{Alyssa Sokol}
\fnmark[3]
\fntext[3]{Present address Dept. of Astronomy, Univ. of Mass. at Amherst, MA, USA}

\author[adrWellesley]{Sarah Maynard}

\author[adrWellesley]{Anna V. Payne}
\fnmark[4]
\fntext[4]{Present address Institute for Astronomy, Univ. of Hawaii, Honolulu, HI, USA}

\author[adrWellesley]{Arden Radford}
\fnmark[5]
\fntext[5]{Present address Dept. of Earth and Environment, Boston Univ., MA, USA}

\author[adrWellesley]{Alessondra Springmann}
\fnmark[6]
\fntext[6]{Present address Southwest Research Institute, Boulder, CO, USA}

\author[adrMIT12]{Richard P. Binzel}

\author[adrUnion]{Francis P. Wilkin}

\author[adrMidkiff]{Emily A. Mailhot}
\fnmark[7]
\fntext[7]{Present address Steward Observatory, Univ. of Arizona, Tucson, AZ, USA}

\author[adrMidkiff]{Alan H. Midkiff}

\author[adrDudley]{April Russell}
\fnmark[8,9]
\fntext[8]{Present address John Wiley \& Sons, Inc., Hoboken, NJ, USA}
\fntext[9]{Present address Concordia Univ., St. Paul, MN, USA}

\author[adrStephens]{Robert D. Stephens}

\author[adrBbO]{Vincent Gardiner}

\author[adrPROMPT]{Daniel E. Reichart}

\author[adrPROMPT]{Joshua Haislip}

\author[adrPROMPT]{Aaron LaCluyze}
\fnmark[10]
\fntext[10]{Present address Dept. of Physics, Central Michigan Univ., Mt. Pleasant, MI, USA}

\author[adrBehrend]{Raoul Behrend}

\author[adrRoy]{Ren{\'{e}} Roy}

\affiliation[adrMIT12]{
organization={Department of Earth, Atmospheric, and Planetary Sciences, Massachusetts Institute of Technology, Rm. 54-424},
addressline={77 Massachusetts Avenue},
city={Cambridge},
state={MA},
statesep={},
postcode={02139},
country={USA}}

\affiliation[adrWellesley]{
organization={Department of Astronomy, Whitin Observatory, Wellesley College},
addressline={106 Central Street},
city={Wellesley},
state={MA},
statesep={},
postcode={02481},
country={USA}}

\affiliation[adrWilliams]{
organization={Department of Astronomy, Williams College},
addressline={880 Main Street},
city={Williamstown},
state={MA},
statesep={},
postcode={01267},
country={USA}}

\affiliation[adrColgate]{
organization={Department of Physics and Astronomy, Colgate University},
addressline={13 Oak Drive},
city={Hamilton},
state={NY},
statesep={},
postcode={13346},
country={USA}}

\affiliation[adrUnion]{
organization={Department of Physics and Astronomy, Union College},
addressline={807 Union Street},
city={Schenectady},
state={NY},
statesep={},
postcode={12308},
country={USA}}

\affiliation[adrMidkiff]{
organization={Star View Hill Education Center},
addressline={120 Wasigan Road},
city={Blairstown},
state={NJ},
statesep={},
postcode={07825},
country={USA}}

\affiliation[adrDudley]{
organization={Dudley Observatory},
addressline={107 Nott Terrace, Suite 201},
city={Schenectady},
state={NY},
statesep={},
postcode={12308},
country={USA}}

\affiliation[adrStephens]{
organization={Santana Observatory},
addressline={11355 Mount Johnson Court},
city={Rancho Cucamonga},
state={CA},
statesep={},
postcode={91737},
country={USA}}

\affiliation[adrBbO]{
organization={Boambee Observatory},
state={NSW},
statesep={},
postcode={2452},
country={Australia}}

\affiliation[adrPROMPT]{
organization={Department of Physics and Astronomy, University of North Carolina},
addressline={120 East Cameron Avenue},
city={Chapel Hill},
state={NC},
statesep={},
postcode={27599},
country={USA}}

\affiliation[adrBehrend]{
organization={Geneva Observatory},
postcode={CH-1290},
postcodesep={},
city={Sauverny},
country={Switzerland}}

\affiliation[adrRoy]{
organization={Blauvac Observatory},
addressline={293 Chemin de St Guillaume},
postcode={84570},
postcodesep={},
city={Blauvac},
country={France}}

\begin{abstract}
%
%
An observational study of Koronis family members' spin properties was
undertaken with two primary objectives:
to reduce selection biases for object
rotation period and lightcurve amplitude
in the sample of members' known spin vectors,
and to better constrain future modeling of spin properties evolution.
%
%
Here we report
rotation lightcurves of nineteen Koronis family members,
and derived results that
increase the sample of determined spin vectors in the Koronis family
to include 34 of the largest 36 family members,
completing it
to $H \approx 11.3$ ($D\sim16$~km)
for the largest 32 members.
The program observations were made
during a total of 72 apparitions between 2005--2021,
and are reported here along with
several earlier unpublished lightcurves.
All of the reported data
were analyzed together with previously published lightcurves
to determine the objects' sidereal rotation periods,
spin vector orientations, and convex model shape solutions.
%
%
The derived distributions of retrograde rotation rates and pole obliquities
appear to be qualitatively consistent with outcomes of modification by
thermal YORP torques.
The distribution of spin rates for the prograde rotators remains
narrower than that for the retrograde rotators;
in particular,
the absence of prograde rotators having periods longer than about 20 h
is real,
while among the retrograde rotators are several objects having longer
periods up to about 65 h.
None of the prograde objects newly added to the sample appear to be
trapped in an $s_6$ spin-orbit resonance
that is characteristic of most of the largest prograde objects
(Vokrouhlick\'{y} et al., 2003);
these smaller objects either
could have been trapped previously and have already evolved out,
or have experienced spin evolution tracks that
did not include the resonance.
\end{abstract}

\begin{highlights}
\item
Koronis family spin vector sample completed to largest 32 members
($\sim 16$ km diameter)
\item
Removal of sample selection biases for interpreting and modeling
spin evolution
\item
The absence of prograde spins having rotation periods longer than about
20 h is real
\item
Smaller prograde objects avoided or escaped the spin-orbit trapping of
larger objects
\item
Epochs analysis combined with convex inversion is effective to
determine spin vectors
\end{highlights}


\maketitle

\section{Introduction}
\label{INTR-SEC}

Studies of spin vector orientations among members of asteroid families
provide valuable information to constrain models of asteroid spin
evolution processes.
Members of a family comprise a group of asteroids that were formed
together from the outcome of collisional disruption of a parent body,
sharing the same physical structure as reaccumulated gravitational
aggregates,
the same age since formation,
and similar evolution for similar lengths of time.
Thus family members serve as a constrained sample that avoids
difficulties of interpreting a distribution of spin properties
which does not share a common origin and dynamical history.

Prior to
determinations of thirteen Koronis family members' spin vectors by
\citet{SLIV03,SLIV09},
spin vectors of main belt asteroids
smaller than about 40~km had not been much explored,
nor had there been
a systematic program of spin vector determinations within
any particular asteroid family,
although statistically larger lightcurve amplitudes of Koronis family
asteroids gave an early indication that Koronis family spin vectors
might have some preferential alignment \citep{BINZ88,BINZ90}.
The markedly non-random
distribution of measured spin vectors determined
initially from observations
of ten of the largest Koronis family members \citep{SLIV02}
revealed evidence of
modification of their spins by YORP
thermal radiation torques,
in some cases combined with solar and planetary gravitational torques
\citep{VOKR03}.
The distribution of spin vectors
among the previously studied sample
is dominated by two groupings:
a group of low-obliquity
retrograde spins spanning a wide range of periods between about 3 h and 30 h,
and
a smaller group of prograde spins with
obliquities near 45$\degsym$ and periods near 8~h,
characteristic of trapping in the $s_6$ spin-orbit resonance
\citep{VOKR03}.
A single ``stray'' longer-period, low-obliquity prograde object
\citep{SLIV09}
subsequently was joined by the largest family member (208)~Lacrimosa
\citep{VOKR21}
which initially had been misidentified as being in the retrograde group
based on an alias sidereal period.

A limitation of the previously studied Koronis family spin vector sample
comprising 13 of the largest 18 family members,
as classified by \citet{MOTH05}
and using relative catalog absolute magnitudes \filter{H}
from the coincident compilation by \citet{THOL09}
as a proxy for relative sizes,
is that selection biases render that sample complete to only the largest 8
members.
Although that set of spin vectors
was sufficient to have revealed evidence of YORP modification in the
distribution of spin rates and spin obliquities,
the small size of the completed sample and the presence of bias effects
limit
its usefulness for further understanding of the spin
evolution.
In particular,
because the rate at which YORP acts on a body is size-dependent,
one may expect systematic study of smaller members of the family
to reveal information that is not apparent
from the distribution of
spin vectors in the incomplete sample of larger members.

\citet{KRYS07} noted the need for more spin vector data for family
main belt asteroids;
subsequently \citet{KRYS13} studied spin vectors of 18 members of the
Flora family and reported evidence of YORP modification,
identifying at least three objects which may be trapped in a
so-called ``Slivan state''
spin-orbit resonance as described by \citet{VOKR03}.
\citet{KIM14} studied spin vectors of 13 members of the Maria family
and did not detect spin vector alignment,
but did find evidence of YORP modification of spin rates
in a sample of 92 members.
\citet{HANU13b} investigated anisotropy of spin vectors in eight
asteroid families including Koronis, Flora, and Maria,
corroborating finding depopulation of poles close to the ecliptic
plane,
but did not find any
candidates
for spin-orbit resonance trapping
in
their 38-member sample from
the Flora family.
In each of the studies
the families' spin vector samples were not controlled for completeness
to any particular size.

Obtaining the data needed for a deliberate spin vector determination program
represents a considerable observing effort
necessarily spanning a number of years.
Determination of a spin vector from ground-based lightcurve
observations requires
that the data set
(a) includes observations from a sufficient number of different viewing aspects,
(b) spans an interval of time long enough to precisely determine the
object's sidereal rotation period,
and (c) samples a sufficient progression of shorter time intervals
within the long time span to rule out alias sidereal periods.
Given that the viewing geometry of a Koronis family member changes very
little during any particular apparition,
observations from at least five to six apparitions are needed to
assemble a suitable lightcurve data set.
\citet{SLIV08a} have described
a long-term observing program designed and undertaken specifically to
obtain the rotation lightcurve data
needed to increase
the sample of determined Koronis family spin vectors;
most of the lightcurves newly reported in the present work
were obtained as part of that program.

The new lightcurves,
like those used to determine the previously-studied sample of
Koronis family spin vectors,
would now be characterized as densely-sampled in time
in contrast with
sparsely-sampled lightcurve data from sky imaging surveys.
Several years into the multi-year observing program,
\citet{DURE09} independently described a method to exploit the
increasing availability of
sparse-in-time lightcurves
by combining them with a smaller number of
densely-sampled lightcurves,
provided that the sparse data brightnesses have been calibrated to some
common zero-point so that they can be assembled
into lightcurves.
The availability and quality
of sparse data to construct a specific sample of any particular asteroids of
interest remain subject to the observing parameters of the sky surveys,
including cadence and targeted brightness range.
Applications of this ``sparse data'' approach \citep{HANU11,HANU13b,
DURE16,DURE18a,DURE18b,DURE19,DURE20}
have yielded spin vector determinations for fifteen of
the observing program objects;
a comparison of those results
with spin vectors determined in the present work for the same objects,
using the same underlying convex modeling approach and assumptions,
but based instead on independent sets of
densely-sampled input lightcurves,
is included in Sec.~\ref{DISC-SEC}.

In this paper the lightcurve observations
and analyses for spins and convex shape models
of nineteen Koronis family members are reported,
completing the sample of determined spin vectors in the Koronis family
down to $H \approx 11.3$ ($D\sim16$~km).
The results
reduce selection biases in the set of known spin vectors
for family members,
and will constrain future modeling of spin properties evolution;
in the meantime,
the spin properties in the expanded sample are briefly discussed.

Catalog physical properties for
the observing program objects are summarized in Table~{\ref{PHYS-TBL}},
which lists for each object
its catalog absolute magnitude {\filter{H}},
available taxonomic classifications, and
whether a spectrum has been reported.
Taxonomic types have been determined for seventeen of the objects,
all of which are classified as either some variety of S-type,
or for (1100)~Arnica
as a union set of two types that includes S-type.

\begin{table*}[width=.9\textwidth,cols=2,pos=h]
\caption{Catalog physical properties for the observing program objects.}
\label{PHYS-TBL}
\begin{tabular*}{\tblwidth}{@{}lrlll@{}}
\toprule
Asteroid&Catalog \filter{H}&Taxonomic        &Spectrum &Taxonomy \\
        &(notes a,b)  &classification(s)&reported?&reference\\
\midrule
  (658) Asteria       &10.51& S         &-- & \citet{NEES10}\\
  (761) Brendelia     &10.74& S, SC     &-- & \citet{NEES10,HASS12}\\
  (811) Nauheima      &10.77& S, S0, LS &-- & \citet{NEES10,HASS12}\\
  (975) Perseverantia &10.50& S         &-- & \citet{NEES10}\\
 (1029) La Plata      &10.92& S         &-- & \citet{NEES10,HASS12}\\
 (1079) Mimosa        &11.19& S         &-- & \citet{NEES10,HASS12}\\
 (1100) Arnica        &10.94& LS        &-- & \citet{HASS12}\\
 (1245) Calvinia      & 9.98& S, S0     &-- & \citet{NEES10}\\
 (1336) Zeelandia     &10.67& S, S0     &yes& \citet{NEES10}\\
 (1350) Rosselia      &10.73& S, Sa     &yes& \citet{NEES10}\\
 (1423) Jose          &10.97& S         &yes& \citet{NEES10,HASS12}\\
 (1482) Sebastiana    &11.10& --        &-- & \\
 (1618) Dawn          &11.23& S         &yes& \citet{NEES10}\\
 (1635) Bohrmann      &11.08& S         &yes& \citet{NEES10}\\
 (1725) CrAO          &11.37& S         &yes& \citet{NEES10}\\
 (1742) Schaifers     &11.32& S         &-- & \citet{HASS12}\\
 (1848) Delvaux       &11.11& S         &yes& \citet{NEES10}\\
 (2144) Marietta      &11.35& S         &-- & \citet{HASS12}\\
 (2209) Tianjin       &11.26& --        &-- & \\
\bottomrule
\end{tabular*}
\begin{flushleft}
\noteindex{a}
Catalog \filter{H} values
published in various Minor Planets and Comets Orbit Supplements
from the IAU Minor Planet Center,
and curated by the Solar System Dynamics group
at the NASA Jet Propulsion Laboratory.
Retrieved using the JPL Small-Body Database Lookup Web tool
\verb#https://ssd.jpl.nasa.gov/tools/sbdb_lookup.html#
on 2022 March 28.
\newline
{\noteindex{b}}
Use of catalog {\filter{H}} in the present work is limited to
serving as proxy for approximate relative sizes of the family members.
Solar phase parameters derived from observing program data
are reported in Table~{\ref{PC-RESULTS-TBL}}.
\end{flushleft}
\end{table*}

\section{Observations}
\label{OBS-SEC}

The observing program
used CCD imaging cameras at
nine different observatories
from 2005 September
through 2021 August,
recording
443 individual rotation lightcurves of the
nineteen target objects
during 72 different apparitions.
Once observations of a target were begun during an apparition,
particular efforts were made to obtain complete coverage
in rotation phase at a sampling rate of at least 50 points per rotation
in order to maximize usefulness of the data for spin vector analyses,
and obtaining data to also determine a more precise rotation period
was prioritized
for objects whose existing period determination was
not precise enough to meet the criterion described by
\citet{SLIV12b}.
During most apparitions, observations were made to calibrate the
lightcurves to standard magnitudes
by observing solar-colored standard stars from \citet{LAND83,LAND92},
and when possible, additional observations were made to increase solar
phase angle coverage.
Images were processed and
measured using standard
techniques for synthetic aperture photometry.
Also reported here with the observing program data are several
previously unpublished lightcurves from one earlier apparition.

Observing circumstances for each object are presented in
Table~\ref{SUMMARYCIRC-TBL},
summarized by apparition.
The table
lists for each apparition
the UT date month(s) during which the observations were made,
the number of individual lightcurves observed,
the approximate ecliptic longitude $\lambda_{\rm PAB}$
and latitude $\beta_{\rm PAB}$ (J2000) of the phase angle bisector,
the range of solar phase angles $\alpha$ observed,
and the filter(s) and telescope used.
Information about the telescopes, observatories, and detectors
is presented in
Table~\ref{TEL-TBL}.
Rotation period results are presented in
Table~\ref{LC-RESULTS-TBL},
which lists for each object
the observed range of peak-to-peak lightcurve amplitudes
and the derived periods with their uncertainties.
For periods improved during this study the apparition whose
lightcurves yield the most precise result is identified.
Color index results
are summarized in
Table~\ref{CI-RESULTS-TBL}.

\begin{table*}[width=.9\textwidth,cols=2,pos=h]
\caption{Summarized observing circumstances.}
\label{SUMMARYCIRC-TBL}
\begin{tabular*}{\tblwidth}{@{}llrrrrll@{}}
  \toprule
Object &
UT date(s) &
\multicolumn{1}{c}{$N_{\rm lc}$} &
\multicolumn{1}{c}{$\lambda_{\rm PAB}$} &
\multicolumn{1}{c}{$\beta_{\rm PAB}$}
&
\multicolumn{1}{c}{$\alpha$} &
Filter(s)\noteindex{a} &
Telescope

\\ \midrule
(658) Asteria       & 2007 Sep           &  4 & 340$\degsym$ &  0$\degsym$ &  0$\degsym$--4$\degsym $ & \filter{V} & b \\[0.25pc]

(761) Brendelia     & 2005 Sep--Nov      & 21 & 353$\degsym$ & $-$1$\degsym$ &  2$\degsym$--18$\degsym$ & \filter{V,R} & b \\
                    & 2006 Dec--2007 Feb & 21 &  90$\degsym$ & +2$\degsym$ &  2$\degsym$--16$\degsym$ & \filter{V} & b \\
                    & 2008 Jan--Apr      & 19 & 170$\degsym$ & +1$\degsym$ &  1$\degsym$--16$\degsym$ & \filter{V,R} & b \\
                    & 2009 May--Jun      &  9 & 260$\degsym$ & $-$2$\degsym$ &  3$\degsym$--9$\degsym $ & \filter{V} & b,c \\
                    & 2013 Mar--May      & 14 & 183$\degsym$ & +1$\degsym$ &  3$\degsym$--17$\degsym$ & \filter{R} & b \\[0.25pc]

(811) Nauheima      & 2005 Dec--2006 Feb &  3 &  98$\degsym$ & $-$2$\degsym$ &  5$\degsym$--17$\degsym$ & \filter{V} & b \\
                    & 2007 Feb--Mar      &  2 & 174$\degsym$ & +3$\degsym$ &  3$\degsym$--7$\degsym $ & \filter{V} & b \\[0.25pc]

(975) Perseverantia & 2005 Sep           &  1 & 357$\degsym$ & $-$2$\degsym$ &  6$\degsym$              & \filter{V} & b \\
                    & 2006 Nov--2007 Feb &  6 &  97$\degsym$ & +3$\degsym$ &  6$\degsym$--19$\degsym$ & \filter{V,R} & b \\
                    & 2008 Mar--Apr      &  4 & 197$\degsym$ & +1$\degsym$ &  3$\degsym$--9$\degsym $ & \filter{V} & b \\
                    & 2013 May           &  5 & 218$\degsym$ &  0$\degsym$ &  2$\degsym$--6$\degsym $ & \filter{R} & b \\[0.25pc]

(1029) La Plata     & 2005 Dec           &  4 &  78$\degsym$ & +2$\degsym$ &  1$\degsym$--3$\degsym $ & \filter{V} & b \\
                    & 2007 Mar--Apr      &  4 & 173$\degsym$ & +2$\degsym$ &  1$\degsym$--15$\degsym$ & \filter{V} & b \\[0.25pc]

(1079) Mimosa       & 1984 Nov           &  5 &  68$\degsym$ & +2$\degsym$ &  1$\degsym$--4$\degsym $ & \filter{B} & k \\
                    & 2013 Oct--Nov      & 14 &  43$\degsym$ & +1$\degsym$ &  1$\degsym$--13$\degsym$ & \filter{R} & b \\
                    & 2016 Apr--May      &  8 & 235$\degsym$ & $-$1$\degsym$ &  1$\degsym$--7$\degsym $ & \filter{R} & l \\
                    & 2017 Jul--Aug      & 11 & 319$\degsym$ &  0$\degsym$ &  2$\degsym$--8$\degsym $ & \filter{R} & l \\
                    & 2021 May--Jul      & 30 & 247$\degsym$ & $-$1$\degsym$ &  6$\degsym$--18$\degsym$ & \filter{r',R} & m,n \\[0.25pc]

(1100) Arnica       & 2007 Sep--Nov      &  7 &  20$\degsym$ & +1$\degsym$ &  1$\degsym$--12$\degsym$ & \filter{V,R} & b \\
                    & 2009 Jan--Mar      &  9 & 111$\degsym$ &  0$\degsym$ &  3$\degsym$--18$\degsym$ & \filter{V,R} & b \\
                    & 2010 Feb--Apr      &  7 & 188$\degsym$ & $-$1$\degsym$ &  2$\degsym$--12$\degsym$ & c          & d \\
                    & 2011 May--Jun      &  6 & 279$\degsym$ & $-$1$\degsym$ &  3$\degsym$--14$\degsym$ & \filter{V} & c,e \\
                    & 2012 Sep--2013 Jan &  6 &  32$\degsym$ & +1$\degsym$ &  1$\degsym$--20$\degsym$ & \filter{V,R} & b \\
                    & 2014 Jan           &  5 & 115$\degsym$ &  0$\degsym$ &  2$\degsym$--6$\degsym $ & \filter{R} & b \\[0.25pc]

(1245) Calvinia     & 2006 Jul           &  3 & 269$\degsym$ & +3$\degsym$ &  9$\degsym$--13$\degsym$ & \filter{V} & b \\
                    & 2007 Oct           &  1 &  15$\degsym$ & $-$3$\degsym$ &  4$\degsym$              & \filter{V} & b \\
                    & 2012 Dec           &  1 &  26$\degsym$ & $-$3$\degsym$ & 17$\degsym$              & \filter{V} & b \\[0.25pc]

(1336) Zeelandia    & 2006 Oct           &  5 & 335$\degsym$ & $-$4$\degsym$ & 14$\degsym$--20$\degsym$ & \filter{V,R} & b \\
                    & 2007 Nov-Dec       &  2 &  72$\degsym$ & $-$2$\degsym$ &  2$\degsym$              & \filter{V} & b \\
                    & 2009 Mar--Apr      &  3 & 155$\degsym$ & +3$\degsym$ &  7$\degsym$--17$\degsym$ & \filter{V,R} & b \\
                    & 2010 Jun           &  5 & 245$\degsym$ & +2$\degsym$ &  4$\degsym$--7$\degsym $ & \filter{V} & c \\
                    & 2011 Aug--Oct      &  6 & 352$\degsym$ & $-$4$\degsym$ &  8$\degsym$--14$\degsym$ & \filter{V} & e \\
                    & 2013 Jan--Mar      &  4 &  91$\degsym$ &  0$\degsym$ &  8$\degsym$--19$\degsym$ & \filter{R} & b \\
                    & 2014 Mar--Apr      &  6 & 170$\degsym$ & +4$\degsym$ &  7$\degsym$--14$\degsym$ & \filter{R} & b \\[0.25pc]

(1350) Rosselia     & 2006 Sep           &  3 & 335$\degsym$ & $-$1$\degsym$ &  6$\degsym$--10$\degsym$ & \filter{V,R} & b \\
                    & 2007 Dec           &  2 &  89$\degsym$ & $-$3$\degsym$ &  1$\degsym$--2$\degsym $ & \filter{V} & b \\[0.25pc]

(1423) Jose         & 2006 Jan--Mar      &  6 & 139$\degsym$ & +4$\degsym$ &  4$\degsym$--14$\degsym$ & \filter{V} & b \\
                    & 2007 Apr--Jun      &  7 & 218$\degsym$ & +1$\degsym$ &  2$\degsym$--14$\degsym$ & \filter{V} & b,f \\[0.25pc]

(1482) Sebastiana   & 2006 Mar--Apr      &  4 & 151$\degsym$ &  +$4\degsym$ & 5$\degsym$--18$\degsym$ & \filter{V} & b \\
                    & 2007 May--Jun      &  4 & 244$\degsym$ &   $0\degsym$ & 1$\degsym$--9$\degsym $ & \filter{V} & b \\[0.25pc]

(1618) Dawn         & 2014 Jan           &  5 &  99$\degsym$ &  0$\degsym$ &  3$\degsym$--12$\degsym$ & \filter{R} & b \\
                    & 2015 Feb--Apr      & 11 & 187$\degsym$ & +4$\degsym$ &  2$\degsym$--15$\degsym$ & \filter{R} & b \\
                    & 2017 Oct--Dec      &  9 &  25$\degsym$ & $-$4$\degsym$ &  2$\degsym$--18$\degsym$ & \filter{R} & b \\
                    & 2020 Jun           &  4 & 205$\degsym$ & +3$\degsym$ & 20$\degsym$              & \filter{R} & i \\
                    & 2021 Jun--Aug      & 12 & 299$\degsym$ & $-$1$\degsym$ &  3$\degsym$--15$\degsym$ & \filter{R} & m \\[0.25pc]

\bottomrule
\end{tabular*}
\end{table*}

\begin{table*}[width=.9\textwidth,cols=2,pos=h]
{\normalsize Table~\ref{SUMMARYCIRC-TBL} (\em{continued})}
\begin{tabular*}{\tblwidth}{@{}llrrrrll@{}}
  \toprule
Object &
UT date(s) &
\multicolumn{1}{c}{$N_{\rm lc}$} &
\multicolumn{1}{c}{$\lambda_{\rm PAB}$} &
\multicolumn{1}{c}{$\beta_{\rm PAB}$}
&
\multicolumn{1}{c}{$\alpha$} &
Filter(s)\noteindex{a} &
Telescope
\\ \midrule
(1635) Bohrmann     & 2007 Jul--Aug      &  6 & 273$\degsym$ & +2$\degsym$ & 11$\degsym$--18$\degsym$ & \filter{R} & b \\
                    & 2008 Oct--Nov      & 10 &  20$\degsym$ & $-$1$\degsym$ &  1$\degsym$--14$\degsym$ & \filter{V,R} & b \\
                    & 2009 Nov--2010 Feb &  6 & 110$\degsym$ & $-$2$\degsym$ &  3$\degsym$--15$\degsym$ & \filter{V,R} & b,d,g \\
                    & 2011 Apr--May      &  3 & 192$\degsym$ & +1$\degsym$ &  8$\degsym$--11$\degsym$ & c & d \\
                    & 2012 Jun--Jul      &  6 & 290$\degsym$ & +2$\degsym$ &  1$\degsym$--11$\degsym$ & \filter{V,R} & b,c \\[0.25pc]

(1725) CrAO         & 2014 Sep--Oct      &  8 & 353$\degsym$ & $-$3$\degsym$ &  3$\degsym$--16$\degsym$ & \filter{V,r'} & b \\
                    & 2016 Jan--Feb      &  5 &  95$\degsym$ & $-$1$\degsym$ &  7$\degsym$--15$\degsym$ & \filter{V,R} & b \\
                    & 2017 Mar--Apr      &  7 & 171$\degsym$ & +3$\degsym$ &  4$\degsym$--15$\degsym$ & \filter{R} & b \\
                    & 2018 Jul--Aug      &  5 & 253$\degsym$ & +2$\degsym$ & 16$\degsym$--20$\degsym$ & \filter{R} & h \\
                    & 2019 Aug           &  3 & 356$\degsym$ & $-$3$\degsym$ & 10$\degsym$--13$\degsym$ & \filter{R} & h \\[0.25pc]

(1742) Schaifers    & 2007 Jul--Aug      &  6 & 289$\degsym$ & +2$\degsym$ &  2$\degsym$--9$\degsym $ & \filter{R} & b \\
                    & 2013 Oct--2014 Feb & 12 &  51$\degsym$ & $-$3$\degsym$ &  2$\degsym$--21$\degsym$ & \filter{V,R} & b \\
                    & 2015 Feb           &  2 & 142$\degsym$ & $-$1$\degsym$ &  1$\degsym$--4$\degsym $ & \filter{R} & b \\
                    & 2016 Apr--May      &  2 & 217$\degsym$ & +3$\degsym$ &  1$\degsym$--5$\degsym $ & \filter{R} & b \\[0.25pc]

(1848) Delvaux      & 2006 Nov--Dec      &  2 & 336$\degsym$ &  0$\degsym$ & 20$\degsym$              & \filter{R} & b \\
                    & 2009 Mar--Apr      &  2 & 152$\degsym$ &  0$\degsym$ & 10$\degsym$--14$\degsym$ & \filter{V,R} & b \\
                    & 2013 Feb           &  1 &  79$\degsym$ & +2$\degsym$ & 18$\degsym$              & \filter{R} & b \\
                    & 2014 Apr           &  2 & 164$\degsym$ & $-$1$\degsym$ & 15$\degsym$--17$\degsym$ & \filter{R} & b \\
                    & 2015 Jun           &  1 & 259$\degsym$ & $-$2$\degsym$ &  6$\degsym$              & \filter{R} & j \\[0.25pc]

(2144) Marietta     & 2007 Oct--Nov      &  4 &  28$\degsym$ & $-$3$\degsym$ &  2$\degsym$--9$\degsym $ & \filter{V} & b \\
                    & 2015 Mar--Jun      &  3 & 230$\degsym$ & +4$\degsym$ & 10$\degsym$--15$\degsym$ & \filter{R} & b \\
                    & 2016 Aug           &  1 & 313$\degsym$ &  0$\degsym$ &  1$\degsym$              & \filter{R} & b \\[0.25pc]

(2209) Tianjin      & 2007 Nov--Dec      &  4 &  57$\degsym$ & $-$3$\degsym$ &  4$\degsym$--10$\degsym$ & \filter{V} & b \\
                    & 2009 Apr           &  6 & 162$\degsym$ & +1$\degsym$ & 15$\degsym$--18$\degsym$ & \filter{V,R} & b \\
                    & 2013 Feb--Apr      &  5 &  89$\degsym$ & $-$2$\degsym$ & 20$\degsym$--21$\degsym$ & \filter{R} & b \\
                    & 2014 Mar--Apr      &  6 & 178$\degsym$ & +2$\degsym$ &  4$\degsym$--12$\degsym$ & \filter{R} & b \\
                    & 2016 Sep           &  2 & 351$\degsym$ & $-$1$\degsym$ &  4$\degsym$--5$\degsym$ & \filter{R} & b \\[0.25pc]
\bottomrule
\end{tabular*}
\begin{flushleft}
\noteindex{a}
Filters:
\filter{B}, Johnson \filter{B};
\filter{V}, Johnson \filter{V};
\filter{R}, Cousins \filter{R};
\filter{r'}, SDSS \filter{r'};
c, colorless.
\newline
\noteindex{b}
0.6-m Sawyer, Whitin Obs.,
observers S.\ Slivan and
Corps of Loyal Observers, Wellesley Division (CLOWD)
\newline
\noteindex{c}
0.4-m PROMPT, CTIO,
observers S.\ Slivan, and in 2009 M.\ Hosek
\newline
\noteindex{d}
0.35-m, Santana Obs.,
observer R.\ Stephens
\newline
\noteindex{e}
0.32-m GRAS-009, Siding Spring Obs.,
observer A.\ Russell
\newline
\noteindex{f}
0.3-m, Boambee Obs.,
observer V.\ Gardiner
\newline
\noteindex{g}
0.6-m, Hopkins Obs.,
observer M.\ Hosek
\newline
\noteindex{h}
0.6-m Elliot, Wallace Obs.,
observer S.\ Slivan
\newline
\noteindex{i}
0.6-m Elliot, Wallace Obs.,
observer T.\ Brothers
\newline
\noteindex{j}
0.6-m, Star View Hill Education Center,
observers E.\ Mailhot and A.\ Midkiff
\newline
\noteindex{k}
0.91-m, McDonald Obs.,
observer R.\ Binzel
\newline
\noteindex{l}
0.43-m T17, Siding Spring Obs.,
observer S.\ Slivan
\newline
\noteindex{m}
0.36-m C14 \#3, Wallace Obs.,
observers A.\ Colclasure, I.\ Escobedo,
A.\ Henopp, R.\ Knight, A.\ Mitchell
\newline
\noteindex{n}
0.6-m CHI-1, El Sauce Obs.,
observer F.\ Wilkin
\end{flushleft}
\end{table*}

\clearpage

\begin{table*}[width=.9\textwidth,cols=2,pos=h]
\caption{Telescopes, observatories, detectors.}
\label{TEL-TBL}
\begin{tabular*}{\tblwidth}{@{}lllll@{}}
\toprule
Telescope      & Location                                    &Detector field  &Image           & Note \\
               &                                             & of view ($'$)  & resolution     &      \\
               &                                             &                & ($''$/pix)     &      \\
\midrule
0.61-m Sawyer  & Whitin Obs., Wellesley, MA (2005--2009)     & 16 $\times$ 16 & 1.8            & a \\
0.61-m Sawyer  & Whitin Obs., Wellesley, MA (2012--2016 Feb) & 19 $\times$ 19 & 1.2            & a \\
0.61-m Sawyer  & Whitin Obs., Wellesley, MA (2016 Apr--May)  & 16 $\times$ 16 & 0.9            & a \\
0.61-m Sawyer  & Whitin Obs., Wellesley, MA (2016 Aug--2017 Apr) & 19 $\times$ 19 & 1.2            & a \\
0.61-m Sawyer  & Whitin Obs., Wellesley, MA (2017 Oct--Nov)  & 16 $\times$ 16 & 0.9            & a \\
0.61-m Sawyer  & Whitin Obs., Wellesley, MA (2017 Dec)       & 20 $\times$ 20 & 1.2            & a \\
0.4-m PROMPT   & Cerro Tololo Inter-American Obs., Chile     & 10 $\times$ 10 & 0.6            & a \\
0.61-m Elliot  & Wallace Astrophysical Obs., Westford, MA    & 32 $\times$ 32 & 0.9            & a \\
0.36-m C14 \#3 & Wallace Astrophysical Obs., Westford, MA    & 20 $\times$ 20 & 1.2            & a \\
0.35-m         & Santana Obs., Rancho Cucamonga, CA          & 21 $\times$ 21 & 1.2            & b \\
0.43-m T17     & Siding Spring Obs., Australia               & 16 $\times$ 16 & 0.9            & a \\
0.32-m GRAS009 & Siding Spring Obs., Australia               & 27 $\times$ 18 & 0.7            & b \\
0.6-m CHI-1    & El Sauce Obs., Rio Hurtado, Chile           & 32 $\times$ 32 & 1.2            & a \\
0.6-m          & Hopkins Obs., Williamstown, MA              & 21 $\times$ 21 & 1.2            & a \\
0.6-m          & Star View Hill Obs., Blairstown, NJ         & 11 $\times$ 11 & 2.6            & b,c \\
0.3-m          & Boambee Obs., New South Wales, Australia    & 24 $\times$ 16 & 1.9            &     \\
0.91-m         & McDonald Obs., Ft. Davis, TX                & --             & --             & d   \\
\bottomrule
\end{tabular*}
\begin{flushleft}
\noteindex{a}
Observing and data reduction procedures
as previously described by \citet{SLIV08a}.
\newline
\noteindex{b}
Images measured using the ``Canopus'' application
developed by B. Warner, Palmer Divide Obs., Colorado Springs, CO.
\newline
\noteindex{c}
Equipment and procedures described by \citet{MAIL14}.
\newline
\noteindex{d}
Photometric photomultiplier tube detector;
equipment and procedures described by \citet{BINZ87}.
\end{flushleft}
\end{table*}

\begin{table*}[width=.9\textwidth,cols=2,pos=h]
\caption{Lightcurve amplitude and rotation period results.}
\label{LC-RESULTS-TBL}
\begin{tabular*}{\tblwidth}{@{}llD{.}{.}{4}D{.}{.}{4}l@{}}
\toprule
Asteroid              & Amplitude (mag.) & \multicolumn{1}{l}{Period $P$ (h)}&\multicolumn{1}{l}{$\sigma(P)$}& Period apparition or reference\\
\midrule
  (658) Asteria       & 0.19--0.23 &  21.032   &  0.004    &\citet{SLIV08a} \\
  (761) Brendelia     & 0.15--0.36 &  58.00    &  0.02     & 2013 \\
  (811) Nauheima      & 0.03--0.17 &   4.0011  &  0.0003   &\citet{SLIV08a} \\
  (975) Perseverantia & 0.11--0.25 &   7.224   &  0.001    &\citet{SLIV08a} \\
 (1029) La Plata      & 0.15--0.56 &  15.310   &  0.002    &\citet{SLIV08a} \\
 (1079) Mimosa        & 0.06--0.33 &  64.69    &  0.05     & 2021 \\
 (1100) Arnica        & 0.01--0.30 &  14.533   &  0.002    & 2012--2013 \\
 (1245) Calvinia      & 0.18--0.40 &   4.8512  &  0.0003   &\citet{SLIV08a} \\
 (1336) Zeelandia     & 0.01--0.51 &  15.602   &  0.002    &\citet{SLIV08a} \\
 (1350) Rosselia      & 0.43--0.54 &   8.140   &  0.001    &\citet{SLIV08a} \\
 (1423) Jose          & 0.60--0.73 &  12.3125  &  0.0008   & 2006 \\
 (1482) Sebastiana    & 0.50--0.83 &  10.489   &  0.002    &\citet{SLIV08a} \\
 (1618) Dawn          & 0.27--0.78 &  43.204   &  0.012    & 2015 \\
 (1635) Bohrmann      & 0.01--0.51 &   5.8643  &  0.0003   & 2009--2010 \\
 (1725) CrAO          & 0.07--0.26 &  21.475   &  0.006    & 2014 \\
 (1742) Schaifers     & 0.61--0.95 &   8.533   &  0.001    &\citet{SLIV08a} \\
 (1848) Delvaux       & 0.52--0.81 &   3.6391  &  0.0002   & 2014 \\
 (2144) Marietta      & 0.29--0.49 &   5.4889  &  0.0003   & 2015 \\
 (2209) Tianjin       & 0.23--0.56 &   9.4529  &  0.0007   & 2012--2013 \\
\bottomrule
\end{tabular*}
\end{table*}

\begin{table}
\caption{Color index results.}
\label{CI-RESULTS-TBL}
\begin{tabular*}{\tblwidth}{@{}lllll@{}}
\toprule
 Asteroid         & \colorindex{V}{R} & $\sigma($\colorindex{V}{R}$)$
                  &\colorindex{r'}{R} & $\sigma($\colorindex{r'}{R}$)$ \\
\midrule
  (761) Brendelia &  0.454 & 0.017   & -- & -- \\
 (1079) Mimosa    &  --    & --      &  0.244 & 0.006 \\
 (1100) Arnica    &  0.451 & 0.017   & -- & -- \\
 (1336) Zeelandia &  0.432 & 0.023   & -- & -- \\
 (1350) Rosselia  &  0.466 & 0.018   & -- & -- \\
 (1635) Bohrmann  &  0.452 & 0.023   & -- & -- \\
 (1725) CrAO      &  0.462 & 0.027   & -- & -- \\
 (1742) Schaifers &  0.440 & 0.015   & -- & -- \\
 (1848) Delvaux   &  0.452 & 0.020   & -- & -- \\
 (2209) Tianjin   &  0.443 & 0.035   & -- & -- \\
\bottomrule
\end{tabular*}
\end{table}

Solar phase parameters derived from single apparitions using the
approach described by \citet{SLIV08a} are summarized in
Table~\ref{PC-RESULTS-TBL}.
The table lists
for each solar phase fit result
the object and apparition, the span of solar phase angles fitted,
the broadband filter used,
and the corresponding \filter{H} and \filter{G} values,
where parentheses around the latter
indicate that \filter{G} was adopted rather than fitted.
Absolute magnitudes \filter{H} were determined only from apparitions during
which standard-calibrated lightcurve observations were made at phase
angles of $2\degsym$ or smaller.
Fitted values of the slope parameter \filter{G} were determined only
for apparitions whose observations also span at least seven degrees of
phase angle within the approximately linear part of the brightness model,
at phase angles larger than $5\degsym$;
otherwise, a value for \filter{G} was adopted and held fixed
while fitting for the corresponding \filter{H} value.
Adopted \filter{G} values were calculated from existing fitted values
in \citet{SLIV08a} and in the present paper
when at least one such value is available;
otherwise, they are assigned the value $0.23 \pm 0.11$ \citep{LAGE90}.

\begin{table*}[width=.9\textwidth,cols=2,pos=h]
\caption{Solar phase results.}
\label{PC-RESULTS-TBL}
\begin{tabular*}{\tblwidth}{@{}llllllll@{}}
\toprule
           Asteroid &Apparition& Phase angles fitted&Filter &\filter{H} & $\sigma($\filter{H}$)$ &\filter{G} & $\sigma($\filter{G}$)$ \\
\midrule
  (658) Asteria    &2007      & 0$\degsym$--4$\degsym$  & \filter{V} & 10.47 & 0.01 &(0.17)\noteindex{a}&(0.01) \\[0.25pc]

  (761) Brendelia  &2005      & 2$\degsym$--18$\degsym$ & \filter{R} & 10.36 & 0.02 & 0.29 & 0.01 \\
                   &2006--2007& 2$\degsym$--16$\degsym$ & \filter{V} & 10.82 & 0.02 & 0.23 & 0.01 \\
                   &2008      & 1$\degsym$--15$\degsym$ & \filter{V} & 10.83 & 0.02 & 0.27 & 0.01 \\[0.25pc]

 (1029) La Plata   &2005      & 1$\degsym$--3$\degsym$  & \filter{V} & 10.79 & 0.01 &(0.16)\noteindex{a}&(0.03) \\
                   &2007      & 1$\degsym$--15$\degsym$ & \filter{V} & 11.02 & 0.03 & 0.14 & 0.01 \\[0.25pc]

 (1079) Mimosa     &1984      & 1$\degsym$--4$\degsym$  & \filter{B} & 12.11 & 0.03 &(0.24)\noteindex{a}&(0.01) \\
                   &2013      & 1$\degsym$--13$\degsym$ & \filter{R} & 10.88 & 0.03 & 0.25 & 0.02 \\
                   &2016      & 1$\degsym$--7$\degsym$  & \filter{R} & 10.92 & 0.05 &(0.24)\noteindex{a}&(0.01) \\
                   &2017      & 2$\degsym$--8$\degsym$  & \filter{R} & 10.72 & 0.03 &(0.24)\noteindex{a}&(0.01) \\[0.25pc]

 (1100) Arnica     &2007      & 1$\degsym$              & \filter{R} & 10.72 & 0.02 &(0.23)\noteindex{b}&(0.11) \\
                   &2012--2013& 1$\degsym$--2$\degsym$  & \filter{R} & 10.78 & 0.03 &(0.23)\noteindex{b}&(0.11) \\
                   &2014      & 2$\degsym$--4$\degsym$  & \filter{R} & 10.39 & 0.04 &(0.23)\noteindex{b}&(0.11) \\[0.25pc]

 (1336) Zeelandia  &2007      & 2$\degsym$              & \filter{V} & 10.70 & 0.01 &(0.20)\noteindex{a}&(0.01) \\[0.25pc]

 (1350) Rosselia   &2007      & 2$\degsym$              & \filter{V} & 10.78 & 0.03 &(0.23)\noteindex{b}&(0.11) \\[0.25pc]

 (1423) Jose       &2007      & 2$\degsym$--4$\degsym$  & \filter{V} & 11.04 & 0.02 &(0.15)\noteindex{a}&(0.03) \\[0.25pc]

 (1482) Sebastiana &2007      & 1$\degsym$--5$\degsym$  & \filter{V} & 11.15 & 0.02 &(0.20)\noteindex{a}&(0.02) \\[0.25pc]

 (1618) Dawn       &2015      & 2$\degsym$--15$\degsym$ & \filter{R} & 10.88 & 0.04 & 0.22 & 0.02 \\[0.25pc]

 (1635) Bohrmann   &2008      & 1$\degsym$--14$\degsym$ & \filter{V} & 11.02 & 0.02 & 0.29 & 0.01 \\
                   &2012      & 1$\degsym$--2$\degsym$  & \filter{V} & 11.27 & 0.02 &(0.24)\noteindex{a}&(0.07) \\[0.25pc]

 (1742) Schaifers  &2007      & 2$\degsym$--4$\degsym$  & \filter{R} & 10.97 & 0.02 &(0.17)\noteindex{a}&(0.02) \\
                   &2013      & 2$\degsym$--16$\degsym$ & \filter{R} & 10.85 & 0.03 & 0.17 & 0.02 \\
                   &2015      & 1$\degsym$--4$\degsym$  & \filter{R} & 10.87 & 0.01 &(0.17)\noteindex{a}&(0.02) \\
                   &2016      & 1$\degsym$--5$\degsym$  & \filter{R} & 10.89 & 0.01 &(0.17)\noteindex{a}&(0.02) \\[0.25pc]

 (1848) Delvaux&2014\noteindex{c}&1$\degsym$--4$\degsym$& \filter{R} & 10.73 & 0.03 &(0.23)\noteindex{b}&(0.11) \\[0.25pc]

 (2144) Marietta   &2007      & 2$\degsym$--4$\degsym$  & \filter{V} & 11.40 & 0.03 &(0.23)\noteindex{b}&(0.11) \\
                   &2016      & 1$\degsym$              & \filter{R} & 10.90 & 0.02 &(0.23)\noteindex{b}&(0.11) \\
\bottomrule
\end{tabular*}
\begin{flushleft}
\noteindex{a}
Adopted \filter{G} by combining
the fitted values available for this object
from this work
and
from \citet[Table~3]{SLIV08a}.
\newline
\noteindex{b}
Adopted \filter{G} from \citet{LAGE90}.
\newline
\noteindex{c}
(1848) Delvaux lightcurves from \citet{ARRE14}.
\end{flushleft}
\end{table*}

\begin{figure*}
\centering
\includegraphics[width=166mm]{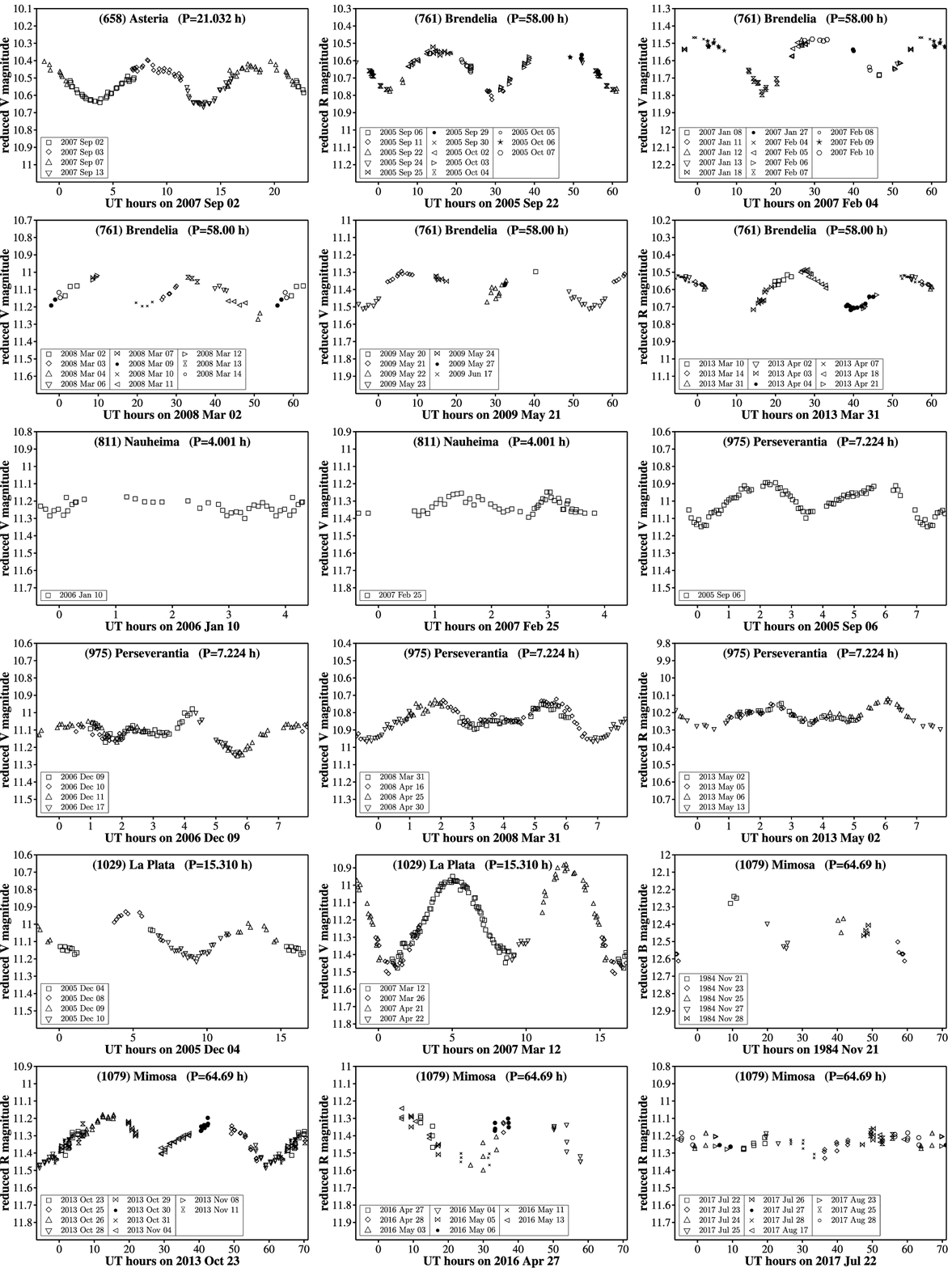}
\caption{Selected composite rotation lightcurves by object and apparition.}
\label{LCFIGSPANEL1-FIG}
\end{figure*}

Selected plots of composite lightcurves from the observing program
are arranged by apparition
in Fig.~\ref{LCFIGSPANEL1-FIG};
the complete data will be made available online.
The time axis on each plot is marked in UT hours on the
composite date,
showing one cycle of the rotation period plus
the earliest and latest 10\% repeated.
Individual nights' data have been
corrected for light-time and folded using the given period;
the legend identifies
the UT dates of the original observations.
Each graph shares
a uniform brightness axis scale spanning 1.0 magnitude.
Standard-calibrated brightnesses have been reduced for changing
distances and solar phase angles;
relative brightnesses were composited by shifting in brightness for
best fit to the composite.

\begin{figure*}
\centering
\includegraphics[width=166mm]{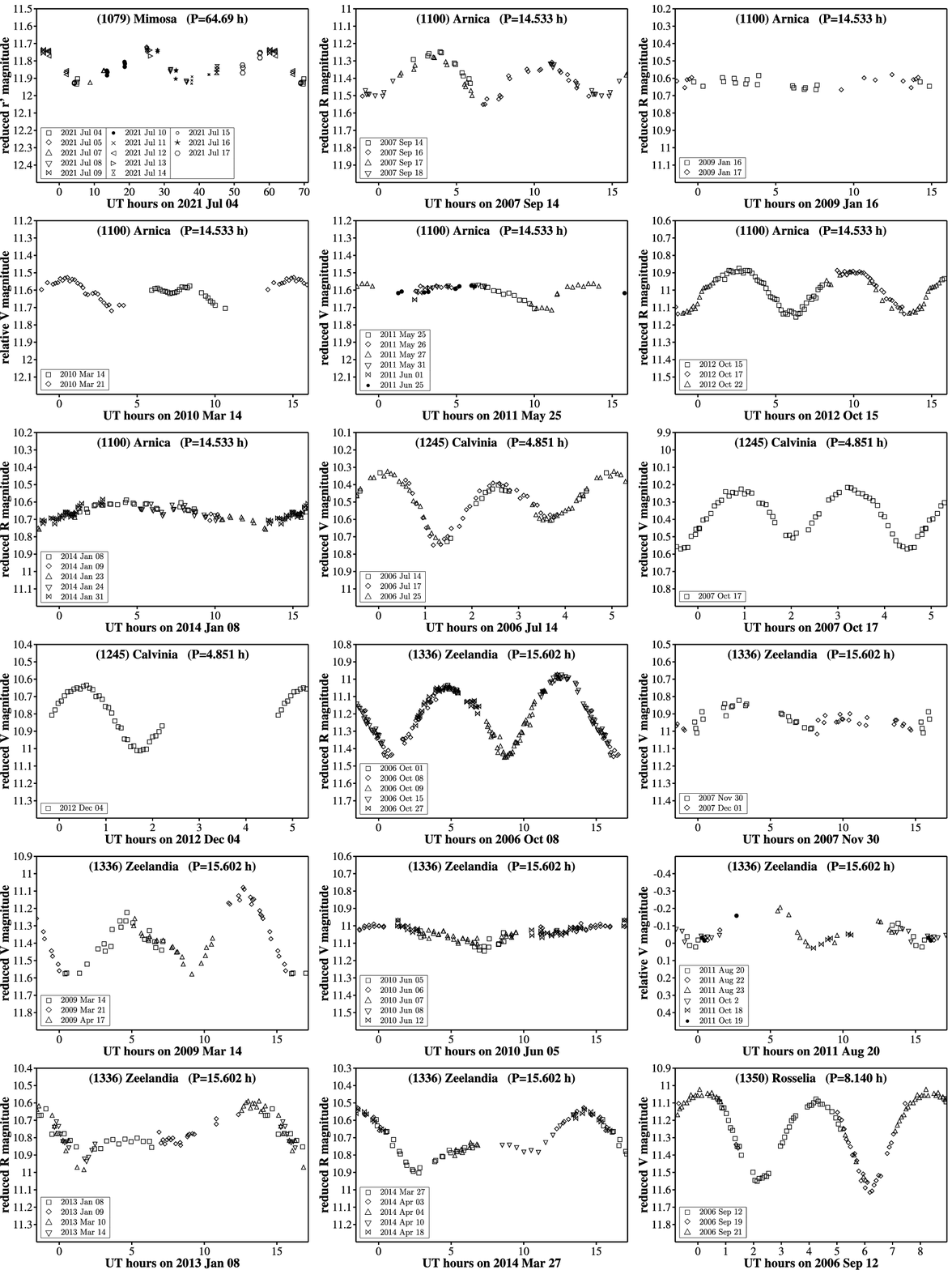}
\center{(Fig.~\ref{LCFIGSPANEL1-FIG} continued)}
\end{figure*}

\begin{figure*}
\centering
\includegraphics[width=166mm]{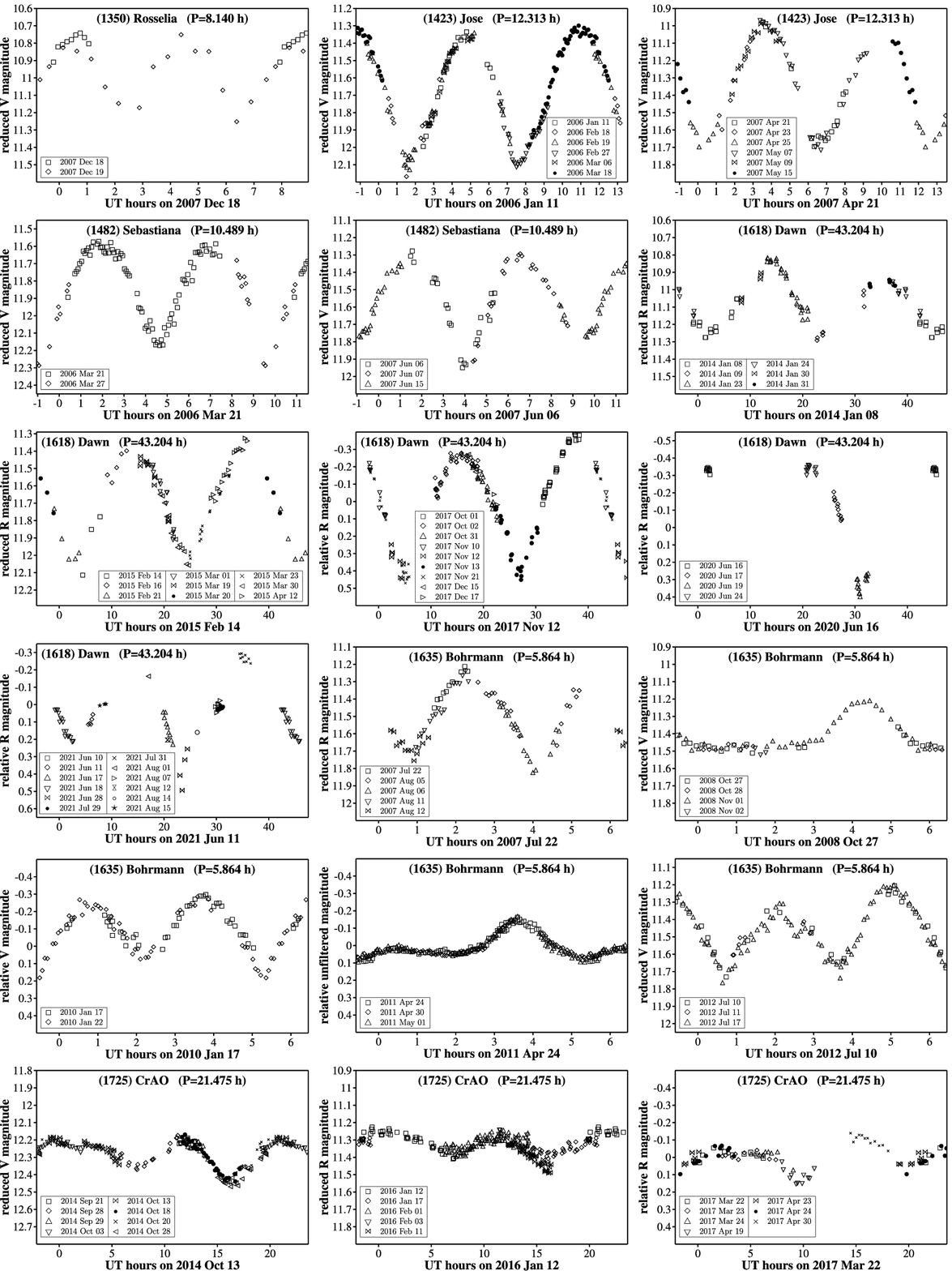}
\center{(Fig.~\ref{LCFIGSPANEL1-FIG} continued)}
\end{figure*}

\begin{figure*}
\centering
\includegraphics[width=166mm]{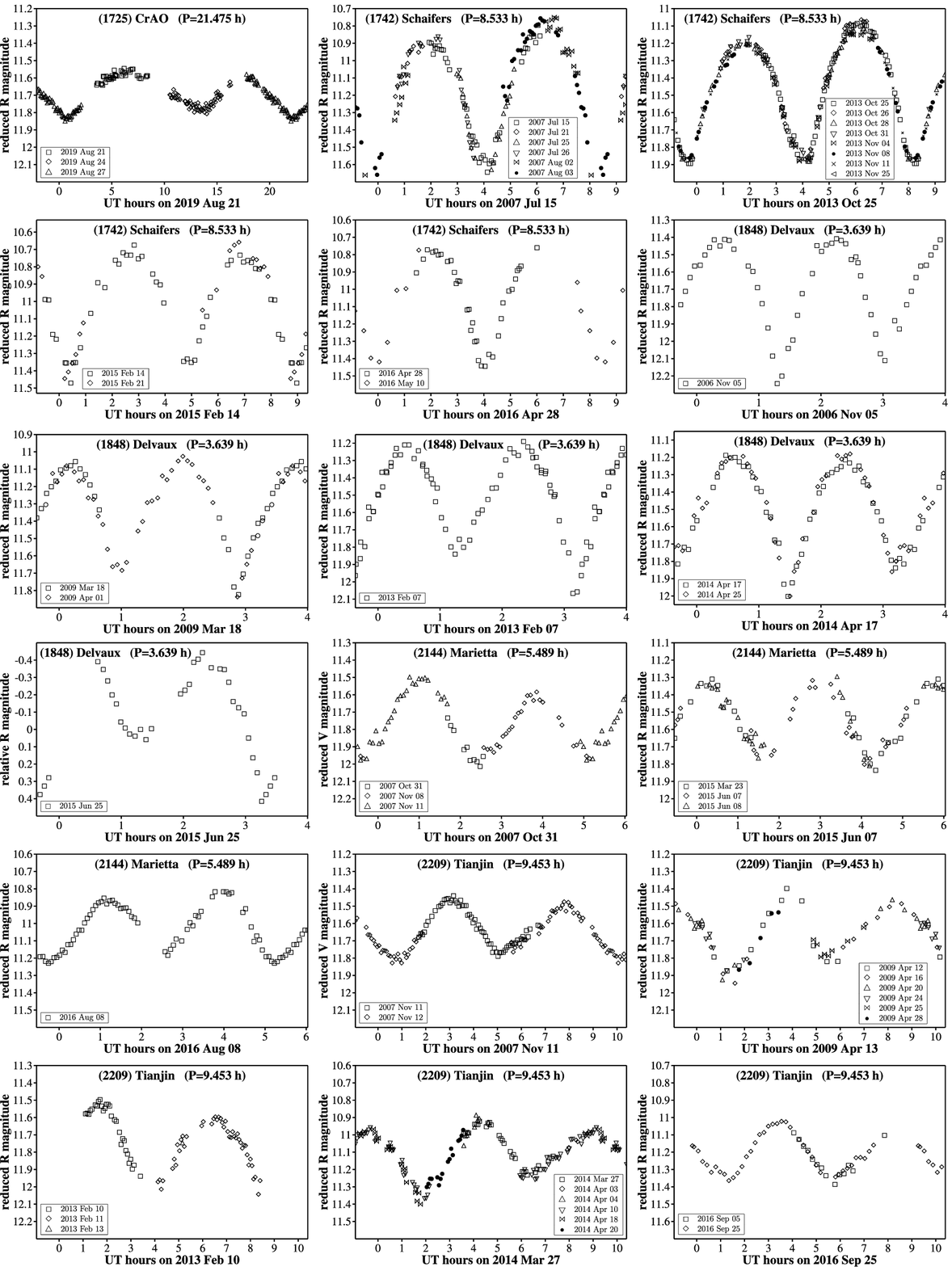}
\center{(Fig.~\ref{LCFIGSPANEL1-FIG} continued)}
\end{figure*}

\clearpage

\begin{figure*}
\centering
\includegraphics[width=166mm]{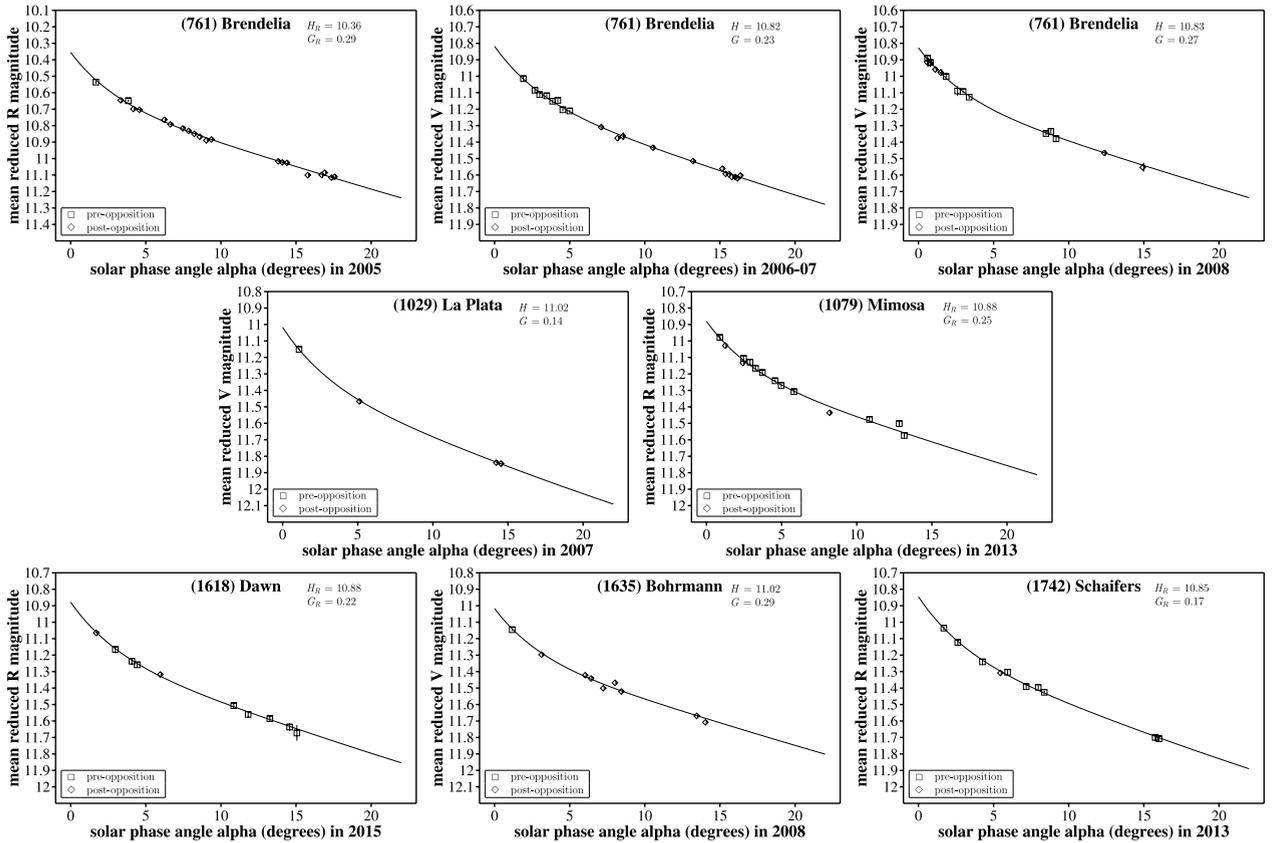}
\caption{Single-apparition solar phase curves.}
\label{PCFIGSPANEL-FIG}
\end{figure*}

Solar phase curves corresponding to
the eight apparitions yielding fitted values for the
slope parameter \filter{G}
are shown in Fig.~\ref{PCFIGSPANEL-FIG},
in which mean reduced magnitudes
are plotted as a function of phase angle.
On each graph the solid curve represents the best fit of the Lumme-Bowell
phase function \citep{BOWE89} to the mean brightnesses;
the corresponding \filter{H}\/ and \filter{G}\/ values appear in the
upper right-hand corner of the plot.

\subsection{Notes on individual objects' lightcurves}

{\em (811) Nauheima.}
The lightcurves recorded during both of the apparitions
reported here exhibit significant shape changes over the periods
of the observations.

{\em (1079) Mimosa.}
A previously unpublished lightcurve
recorded in 1984 during the observing program of \citet{BINZ87}
also is reported here.

{\em (1100) Arnica.}
\citet{SLIV08a}
observed lightcurves from low-amplitude viewing aspects
and initially deduced
a rotation period
that is double the correct value,
and
in a note added in proof
also reported the correct period
based on additional lightcurves that were
recorded in 2007;
those doubly-periodic observations are fully reported here.
The \colorindex{V}{R} color index
reported in Table~\ref{CI-RESULTS-TBL}
is the weighted mean
of independent measurements made
during the 2009 and the 2012--13 apparitions.

{\em (1336) Zeelandia.}
The \colorindex{V}{R} color index
reported in Table~\ref{CI-RESULTS-TBL}
agrees with
the measurement by \citet{THOM11} of $0.431 \pm 0.021$.

{\em (1848) Delvaux.}
\citet{ARRE14} reported lightcurves
from the 2014 apparition
spanning 18 days.
Additional observations made later in the same apparition on two
more nights are reported here,
which extend the observed time span to 61 days 
and were used to determine the synodic rotation period in
Table~\ref{LC-RESULTS-TBL}.

\section{Analysis for sidereal periods, spin vectors, and model shapes}
\label{SV-ANALYSES-SEC}

The spin vector and model shape analyses
were carried out as 
a sequence of five identifiable stages,
as described in this section:
resolving the sidereal rotation count,
determining the sidereal period and direction of spin,
resolving the locations of the
symmetric pair of pole regions,
determining the best-fit spin vectors and a preliminary model shape,
and finally
fitting a refined model shape for presentation.
A comprehensive discussion of the analysis method is presented
as well as are details of its application to particular objects,
not only to
share the basis for
confidence in the final results presented,
but also
as an aid and reference for those who may wish to perform their own
spin vector and model shape analyses of lightcurves,
mindful that only limited such information
has made its way into the literature
since the introduction of the convex inversion method
\citep{KAAS01a,KAAS01b}.

{\em Sidereal period and direction of spin.}
A prerequisite for a correct spin vector analysis is correctly counting
the number of rotations that elapsed during the entire span of
the lightcurve observations,
in order to determine the correct sidereal rotation period.
In that context,
discussion of both the synodic period constraint
and the sampling of a progression of time intervals between epochs
can be found in \citet{SLIV12b,SLIV13}.
Each program object's sidereal period was constrained
based on analysis of epochs that were determined from the lightcurves
composited by apparition,
using a Fourier series filtering approach to
identify information related to low-order object shape.
Most of the program objects' apparitions' lightcurves are
doubly periodic and relatively symmetric with sufficient amplitude to
make measuring the epochs straightforward.
The synodic rotation period from Table~\ref{LC-RESULTS-TBL}
was used to fit a Fourier series model,
choosing an appropriate number of up to eight harmonics based on the
sampling of the data,
then filtered for low-order shape by retaining only the
second harmonic to locate times of lightcurve maxima.
Epoch errors were estimated as
1.5$\times$
the larger of either
the time difference between
the corresponding filtered and unfiltered maxima,
or 1\% of the rotation period
which is on the order of the typical RMS error
of a well-determined fit to epochs.

Composite lightcurves having a gap in coverage longer than
1/4-rotation are
too incomplete to use a doubly-periodic model;
in those cases a singly-periodic model at half the rotation period was
fit instead and then filtered for the fundamental.
In a few cases,
lightcurves having the lowest amplitudes
or the greatest departures from doubly-periodic symmetry
required special attention to locate epochs
and are discussed with the objects' analyses.
Epochs estimated from incomplete or low-amplitude
lightcurves were assigned relatively larger uncertainties.

\begin{figure}
\centering
\includegraphics[scale=1.00]{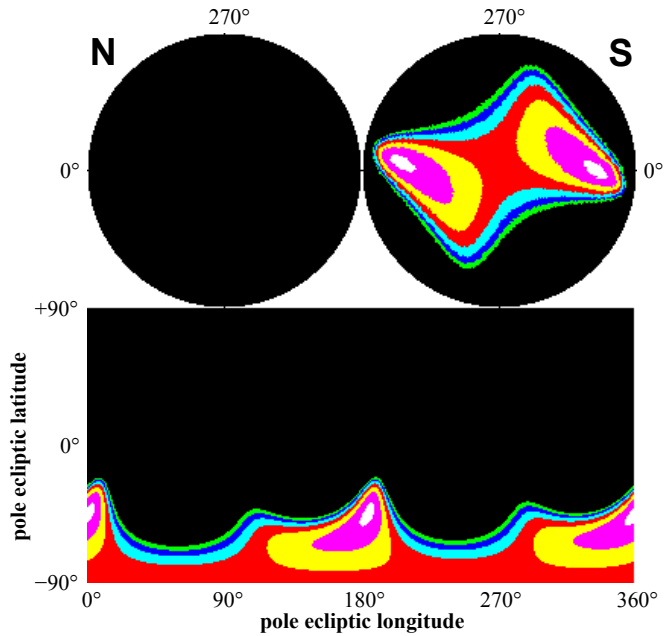}
\caption{
Example contour graph of RMS fit errors using SPA for trial pole
``scans'' for program object (1635) Bohrmann,
showing the entire celestial sphere in ecliptic coordinates
plotted in both polar and rectangular formats,
with areas of best-fit poles colored white.
In this case
the locations of the pair of broader pole regions entirely in
the south ecliptic hemisphere indicate retrograde rotation.
}
\label{A1635-SPASCAN-FIG}
\end{figure}

In the first stage of the epochs analysis for each object,
the ``sieve algorithm'' of
\citet{SLIV13} was used 
with the
synodic rotation periods
in Table~\ref{LC-RESULTS-TBL}
to identify possible sidereal rotation counts,
and their corresponding range(s) of possible sidereal
periods,
that are consistent with the filtered epochs.
For all but one of the program objects the sieve algorithm by itself yielded a
single candidate sidereal rotation count,
leaving only the direction of spin unknown.
Then in the second stage
each candidate period, prograde and retrograde, was checked
as described by \citet{SLIV14},
using the epochs with Sidereal Photometric
Astrometry (SPA) \citep{DRUM88}
to identify the sidereal period, direction of spin,
and a first indication of
either a single pair of symmetric pole solution regions or a
single merged region at an ecliptic pole.
An individual example of pole regions located by SPA
is presented as Fig.~{\ref{A1635-SPASCAN-FIG}}.

The remaining stages of analysis use
the convex inversion (CI) method of \citet{KAAS01b},
a nonlinear iterative approach.
The {\tt convexinv}
implementation\footnote{\tt https://astro.troja.mff.cuni.cz/projects/damit/pages/software\_download}
by J. \v{D}urech was used,
guided by its accompanying documentation
when choosing most of its run-time parameter settings.
An exception is 
that for setting the the zero time $t_0$ and initial
ellipsoid orientation angle $\phi_0$ the suggestion of
\citet{KAAS01b} was followed instead,
choosing values that align the long axis of
the model shape along the $x$-axis.
Doing so avoids very large adjustments to the
model during the first several fit iterations
which can distort the ultimate outcome.
Thus $t_0$ was set to be the time of a brightness minimum from the
filtered Fourier series for a lightcurve recorded at among the smallest
available solar phase angles,
and then $\phi_0$ was calculated as the corresponding pole-dependent
asterocentric longitude of the sub-PAB point
\citep[Eqs. 3--8]{TAYL79,SLIV14}
based on the initial value supplied for the pole location.
The distribution code also was modified to support
user-specified values for the
axial ratios of the triaxial ellipsoid
used as the initial model shape,
and the initial value for $a/b$ generally was
selected based on the maximum amplitude
from among the Fourier series-filtered observed
lightcurves.

The lightcurve data
were binned to a bin width of $1/50$ of the rotation period
for the CI analyses.
Data sets were analyzed as relative photometry,
except for the three longest-rotation period objects
(761)~Brendelia, (1079)~Mimosa, and (1618)~Dawn
whose single-night lightcurves, even from the longest observing nights,
individually do not record enough of a rotation to determine the shape
of the lightcurve without
referencing brightnesses on different nights to a common zero point.
To prepare these objects' data sets for CI,
data whose brightnesses had been
calibrated to standard stars were referenced to a common zero-point
by using available color index information,
and solar phase function coefficients were included as fitted
parameters for CI analysis of these lightcurves as standard-calibrated
photometry.
Lightcurves for these longer-period objects
whose individual nights had been referenced to the same
comparison star but {\em not} also put onto a standard system,
instead were composited in time and in brightness into ``preassembled''
lightcurves of relative photometry.

\begin{figure}
\centering
\includegraphics[scale=1.00]{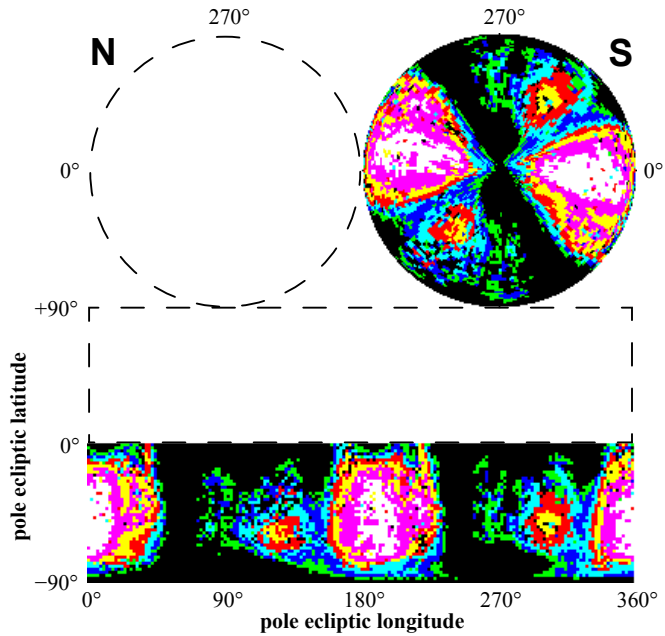}
\caption{
Similar to Fig.~{\protect\ref{A1635-SPASCAN-FIG}} but using CI
and calculated only for trial poles in the southern ecliptic hemisphere.
The pair of pole region locations favored by the lowest $\chi^2$ values
agrees with those of the SPA regions.
}
\label{A1635-CISCAN-FIG}
\end{figure}

{\em Pole regions.}
For the third stage of the analyses,
a contour graph of CI model fits' $\chi^2$ goodness-of-fit
value as a function of
pole location was constructed
to identify the pole regions.
Using the known sidereal period,
a series of CI trials were run
on a grid pattern of trial pole coordinates
having $2\degsym$-resolution in latitude and longitude
to systematically cover
the ecliptic hemisphere indicated by SPA,
in each case holding the pole fixed at the trial location
and iterating until the step change in the RMS deviation
decreased to less than 0.001.
It was found that the pole region location results
were relatively
insensitive to weighting of the lightcurves.
An individual example of pole regions located by CI
is presented as Fig.~{\ref{A1635-CISCAN-FIG}}.

Although pole regions also already are available from the epochs analysis,
the program objects data sets' are large enough 
to expect that the CI region locations will be more reliable,
because SPA pole longitudes can be subject to systematic errors
depending on the degree to which the epochs depart from
the assumption that they all correspond to
the same asterocentric longitude.
The SPA region results did prove useful, however,
to distinguish the correct solutions
when CI allowed more than one possible pair of pole regions.
The remaining symmetric pole ambiguity for the program objects
cannot be resolved using Earth-based lightcurves,
because these asteroids all share low orbital inclinations of 1 to 3 degrees
as members of the Koronis family.
A discussion of symmetry and ambiguities
pertinent to pole solutions obtained from lightcurves
can be found in \citet{MAGN89}.

{\em Spin poles.}
In the fourth stage
CI was run to determine a best-fit period, pole, and preliminary
possible model shape,
using the initial period from SPA
and an initial pole location estimated from the CI contour graph.
For the benefit mainly of the preliminary shape model
a final version of the binned lightcurve
data input was constructed,
weighted typically so that the number of fitted data points per apparition
is proportional to its available rotation phase coverage,
to within a factor of about 1.5.
Any process adjustments needed for a particular object
are described in its individual notes.

For these runs of CI, the criterion for stopping the fit iterations was
given more deliberate consideration than was for the earlier runs to
generate the pole regions contour map.
As an aid in determining at what point to stop the fitting,
the distribution code was modified to add to the fit iteration log at
each step
the fractional change in the $\chi^2$ goodness-of-fit metric,
and the current sidereal period and pole coordinates.
It was found that convergence of the period and pole were straightforward
for these data sets,
but how much longer to keep adjusting the model shape after the period
and pole had converged was not as clear-cut and remained somewhat
subjective;
analyses of certain data sets
showed indications that overfitting could be an issue.
The guideline ultimately adopted was 
to iterate past period and pole convergence
only until the step change
in the fit relative $\chi^2$ decreased to less than about 0.5\%.

Each preliminary model shape result was checked for consistency with
an assumption of of stable rotation about its shortest axis,
by constructing the corresponding model shape rendering and confirming
that its shortest dimension lies along the polar axis.
To solve for the shape model convex polyhedron from its facet areas
and normals,
a C translation of the FORTRAN code for Minkowski minimization from
the CI distribution was used.

The expected symmetry of an object's two ambiguous pole solutions with
respect to the ``photometric great circle'' (PGC) \citep{MAGN89}
was used to perform a self-consistency check on each pair of derived
pole solutions,
as a significant departure from symmetry would indicate the presence of
some problem with the analysis.
Appendix~\ref{APDX-PGC} details how symmetric poles were calculated
using objects' orbit elements.

Errors were estimated
after a solution had passed both consistency checks.
The period error was estimated from SPA with the full set of
Fourier-filtered epochs for a pole at the adopted location.
Uncertainties for the pole locations
were estimated from the CI model fits' $\chi^2$ distributions
in each case by identifying a constant-$\chi^2$ contour
that encloses a region just large enough
to entirely surround
the insignificant fluctuations
near the pole location
within an approximately oval-shaped region.
The contour's angular extents in ecliptic latitude and longitude
were adopted as coarse estimates of $\sim 2\sigma$ confidence,
and each error was rounded to the nearest multiple of $5\degsym$ of arc.
The factor of $\sim 2$ relationship between
the subjective identification of contours and the estimated error
is based on comparison of the published CI pole error estimates
of previously-studied objects
\citep{SLIV03,SLIV09}
with their CI model fits' $\chi^2$ distributions.

{\em Model shape and axial ratios.}
In the last stage of the analysis,
the final shape renderings and axial ratios
corresponding to the best-fit period and pole results
were obtained
using the {\tt conjgradinv} implementation from the CI distribution,
which
uses the facet areas directly rather than using spherical harmonics.
The axial ratios for its initial ellipsoid
were estimated from the preliminary model shape.

\subsection{Spin and shape results}
\label{SPINSHAPE-SEC}

The lightcurves reported in Sec.~\ref{OBS-SEC}
were combined with lightcurves
of the program objects previously published by
\citet{SLIV08a} (46 apparitions)
and by \citet{BINZ87} (6 apparitions),
plus lightcurves from
10 more apparitions whose sources are given
in the analysis descriptions for their specific objects.
Table~\ref{OBS-SUMMARY-TBL}
summarizes for each asteroid's available combined lightcurve data set
the span of years included,
the number of apparitions $N_{\rm ap}$,
and aspect information as a list of ecliptic longitudes $\lambda_{\rm PAB}$
of the
asteroids' phase angle bisector (PAB) near the mid-date of the
observations from each apparition.

\begin{table*}[width=.9\textwidth,cols=2,pos=h]
\caption{Summary of lightcurve observations used for spin vector analyses.}
\label{OBS-SUMMARY-TBL}
\begin{tabular*}{\tblwidth}{@{}llll@{}}
\toprule
  Asteroid        & Years observed  & $N_{\rm ap}$ & $\lambda_{\rm PAB}$ of aspects observed $(\degsym)$ \\
\midrule
  (658) Asteria       & 1983--2007  & 6 &  53,  68, 167, 323, 340, 352 \\
  (761) Brendelia     & 2001--2013  & 7 &  75,  90, 158, 170, 183, 260, 353 \\
  (811) Nauheima      & 1984--2007  & 7 &   9,  98, 174, 239, 253, 340, 347 \\
  (975) Perseverantia & 1990--2013  & 8 &  74,  97, 176, 197, 218, 267, 296, 357 \\
 (1029) La~Plata      & 1975--2007  & 7 &  32,  78, 141, 173, 228, 328, 350 \\
 (1079) Mimosa        & 1983--2021  & 8 &  20,  43,  68, 120, 235, 247, 319, 329 \\
 (1100) Arnica        & 1999--2014  & 9 &  20,  32,  99, 104, 111, 115, 183, 188, 279 \\
 (1245) Calvinia      & 1977--2012  & 7 &   8,  15,  26, 107, 186, 269, 323 \\
 (1336) Zeelandia     & 1999--2014  & 9 &  72,  91, 125, 140, 155, 170, 245, 335, 352 \\
 (1350) Rosselia      & 1975--2007  & 8 &  72,  89, 149, 162, 162, 240, 284, 335 \\
 (1423) Jose          & 2002--2007  & 5 &  41, 139, 206, 218, 290 \\
 (1482) Sebastiana    & 1984--2007  & 7 &  19,  66, 151, 230, 244, 319, 333 \\
 (1618) Dawn          & 2003--2021  & 6 &  25,  75,  99, 187, 205, 299 \\
 (1635) Bohrmann      & 2003--2012  & 7 &   1,  20,  95, 110, 192, 273, 290 \\
 (1725) CrAO          & 2003--2019  & 7 &  95, 171, 236, 253, 341, 353, 356 \\
 (1742) Schaifers     & 1983--2016  & 7 &  32,  51, 129, 142, 217, 289, 350 \\
 (1848) Delvaux       & 2004--2015  & 7 &  79, 140, 152, 164, 259, 336, 349 \\
 (2144) Marietta      & 1999--2016  & 7 &  28, 108, 120, 145, 220, 230, 313 \\
 (2209) Tianjin       & 1996--2016  & 6 &  57,  89, 162, 178, 278, 351 \\
\bottomrule
\end{tabular*}
\end{table*}

The spin vector results for the program objects
are presented in Table~\ref{ALL-SV-RESULTS-TBL}.
The first three columns list the asteroid,
its derived sidereal period $P_{\rm sid}$, and its period error.
The symmetric pair of pole solutions is given in
the next two groups of three columns,
each grouped as a pole identifier followed by
its J2000 ecliptic coordinates $(\lambda_0; \beta_0)$.
For convenient reference
the pole ID notation previously used
by \citet{SLIV03,SLIV09} is retained,
in which \polenum{1} and \polenum{2} denote prograde poles
and \polenum{3} and \polenum{4} denote retrograde poles.
The last two columns in the table give
the estimated uncertainties for both
poles' longitudes and latitudes respectively
in degrees of arc.

\begin{table*}[width=.9\textwidth,cols=2,pos=h]
\caption{Spin vector results.}
\label{ALL-SV-RESULTS-TBL}
\begin{tabular*}{\tblwidth}{@{}lD{.}{.}{8}D{.}{.}{8}crrcrrrr@{}}
  \toprule
  Asteroid&\multicolumn{1}{c}{$P_{\rm sid}$}&\multicolumn{1}{c}{$\sigma(P_{\rm sid})$}& Pole &\multicolumn{1}{c}{$\lambda_0$} &\multicolumn{1}{c}{$\beta_0$} & Pole &\multicolumn{1}{c}{$\lambda_0$} &\multicolumn{1}{c}{$\beta_0$} & $\sigma(\lambda_0)$ & $\sigma(\beta_0)$ \\
&\multicolumn{1}{c}{(h)}&\multicolumn{1}{c}{(h)}&ID&\multicolumn{1}{c}{$(\degsym)$} &\multicolumn{1}{c}{$(\degsym)$} &ID&\multicolumn{1}{c}{$(\degsym)$} &\multicolumn{1}{c}{$(\degsym)$} & \multicolumn{1}{c}{$(\degsym)$}& \multicolumn{1}{c}{$(\degsym)$}\\
\midrule
 (658) Asteria      & 21.02981    & 0.00003     &\polenum{3} &124 &$-$36 &\polenum{4} &306 &$-$39 & 5 & 5 \\
 (761) Brendelia    & 57.9949     & 0.0003      &\polenum{3} &37  &$-$47 &\polenum{4} &212 &$-$48 & 5 & 10 \\
 (811) Nauheima     & 4.000948    & 0.000002    &\polenum{1} &150 &$+$63 &\polenum{2} &343 &$+$60 & 5 & 5 \\
 (975) Perseverantia& 7.223875    & 0.000005    &\polenum{3} &67  &$-$55 &\polenum{4} &238 &$-$58 & 5 & 10 \\
(1029) La~Plata     & 15.310657   & 0.000003    &\polenum{1} &98  &$+$60 &\polenum{2} &281 &$+$55 & 5 & 10 \\
(1079) Mimosa       & 64.72834    & 0.00012     &\polenum{3} &137 &$-$37 &\polenum{4} &318 &$-$39 & 5 & 10 \\
(1100) Arnica       & 14.53191    & 0.00002     &\polenum{1} &123 &$+$28 &\polenum{2} &301 &$+$28 & 5 & 10 \\
(1245) Calvinia     & 4.8514754   & 0.0000005   &\polenum{3} &54  &$-$50 &\polenum{4} &235 &$-$43 & 5 & 5 \\
(1336) Zeelandia    & 15.60188    & 0.00003     &\polenum{1} &45  &$+$5  &\polenum{2} &226 &$+$11 & 5 & 5 \\
(1350) Rosselia     & 8.140103    & 0.000002    &\polenum{3} &131 &$-$84 &\polenum{4} &269 &$-$80 & 5 & 5 \\
(1423) Jose         & 12.312679   & 0.000010    &\polenum{3} &98  &$-$80 &\polenum{4} &237 &$-$82 & 5 & 5 \\
(1482) Sebastiana   & 10.489660   & 0.000003    &\polenum{3} &93  &$-$78 &\polenum{4} &237 &$-$79 & 5 & 5 \\
(1618) Dawn         & 43.22002    & 0.00012     &\polenum{3} &101 &$-$57 &\polenum{4} &268 &$-$58 & 5 & 5 \\
(1635) Bohrmann     & 5.864286    & 0.000003    &\polenum{3} &9   &$-$46 &\polenum{4} &194 &$-$46 & 5 & 10 \\
(1725) CrAO         & 21.47246    & 0.00007     &\polenum{3} &64  &$-$39 &\polenum{4} &241 &$-$33 & 5 & 5 \\
(1742) Schaifers    & 8.532708    & 0.000002    &\polenum{1} &27  &$+$71 &\polenum{2} &195 &$+$75 & 5 & 5 \\
(1848) Delvaux      & 3.63911184  & 0.00000019  &\polenum{3} &143 &$-$83 &\polenum{4} &348 &$-$82 & 5 & 5 \\
(2144) Marietta     & 5.4885454   & 0.0000012   &\polenum{1} &145 &$+$72 &\polenum{2} &344 &$+$70 & 5 & 5 \\
(2209) Tianjin      & 9.452694    & 0.000003    &\polenum{1} &19  &$+$68 &\polenum{2} &186 &$+$72 & 5 & 5 \\
\bottomrule
\end{tabular*}
\end{table*}

The spin obliquity $\varepsilon$ of each pole solution,
calculated as the angle between the spin pole and the orbit pole,
is given in Table~\ref{OBLIQUITIES-TBL}.
In the context of the spin poles,
``prograde'' and ``retrograde'' refer to poles
whose obliquities are less than $90\degsym$
and greater than $90\degsym$, respectively.
``High obliquities'' are near $90\degsym$
and near the orbit plane;
``low obliquities'' are near either $0\degsym$ (prograde)
or $180\degsym$ (retrograde)
and near perpendicular to the orbit plane.
Owing to the small orbital inclinations of the Koronis family,
the obliquities of the two poles comprising
each symmetric pair 
differ by less than $2\degsym$,
well within the estimated uncertainties of the pole locations.
In the last column
the mean of each pair of obliquities is given
as a single adopted mean obliquity for the object.

\begin{table*}[width=.9\textwidth,cols=2,pos=h]
\caption{Spin obliquities
for the spin vectors given in Table~\protect\ref{ALL-SV-RESULTS-TBL}.}
\label{OBLIQUITIES-TBL}
\begin{tabular*}{\tblwidth}{@{}lrlrlrr@{}}
  \toprule
Asteroid & Orbit pole\noteindex{a} & \multicolumn{4}{l}{Spin vector obliquities}  &Adopted $\varepsilon$\\
\midrule
 (658) Asteria       & (261$\degsym$;$+$88$\degsym$) & \polenum{3} & 127$\degsym$ & \polenum{4} & 128$\degsym$ & 127$\degsym$ \\
 (761) Brendelia     & (294$\degsym$;$+$88$\degsym$) & \polenum{3} & 138$\degsym$ & \polenum{4} & 138$\degsym$ & 138$\degsym$ \\
 (811) Nauheima      &  (41$\degsym$;$+$87$\degsym$) & \polenum{1} &  28$\degsym$ & \polenum{2} &  28$\degsym$ &  28$\degsym$ \\
 (975) Perseverantia & (309$\degsym$;$+$87$\degsym$) & \polenum{3} & 146$\degsym$ & \polenum{4} & 147$\degsym$ & 147$\degsym$ \\
(1029) La~Plata      & (300$\degsym$;$+$88$\degsym$) & \polenum{1} &  32$\degsym$ & \polenum{2} &  33$\degsym$ &  32$\degsym$ \\
(1079) Mimosa        & (240$\degsym$;$+$89$\degsym$) & \polenum{3} & 127$\degsym$ & \polenum{4} & 129$\degsym$ & 128$\degsym$ \\
(1100) Arnica        & (214$\degsym$;$+$89$\degsym$) & \polenum{1} &  62$\degsym$ & \polenum{2} &  62$\degsym$ &  62$\degsym$ \\
(1245) Calvinia      &  (62$\degsym$;$+$87$\degsym$) & \polenum{3} & 137$\degsym$ & \polenum{4} & 136$\degsym$ & 136$\degsym$ \\
(1336) Zeelandia     &   (7$\degsym$;$+$87$\degsym$) & \polenum{1} &  83$\degsym$ & \polenum{2} &  82$\degsym$ &  82$\degsym$ \\
(1350) Rosselia      &  (50$\degsym$;$+$87$\degsym$) & \polenum{3} & 172$\degsym$ & \polenum{4} & 172$\degsym$ & 172$\degsym$ \\
(1423) Jose          & (329$\degsym$;$+$87$\degsym$) & \polenum{3} & 171$\degsym$ & \polenum{4} & 172$\degsym$ & 171$\degsym$ \\
(1482) Sebastiana    & (341$\degsym$;$+$87$\degsym$) & \polenum{3} & 169$\degsym$ & \polenum{4} & 169$\degsym$ & 169$\degsym$ \\
(1618) Dawn          &  (13$\degsym$;$+$87$\degsym$) & \polenum{3} & 147$\degsym$ & \polenum{4} & 148$\degsym$ & 148$\degsym$ \\
(1635) Bohrmann      &  (95$\degsym$;$+$88$\degsym$) & \polenum{3} & 136$\degsym$ & \polenum{4} & 136$\degsym$ & 136$\degsym$ \\
(1725) CrAO          &  (29$\degsym$;$+$87$\degsym$) & \polenum{3} & 127$\degsym$ & \polenum{4} & 126$\degsym$ & 126$\degsym$ \\
(1742) Schaifers     &  (62$\degsym$;$+$88$\degsym$) & \polenum{1} &  17$\degsym$ & \polenum{2} &  17$\degsym$ &  17$\degsym$ \\
(1848) Delvaux       & (242$\degsym$;$+$89$\degsym$) & \polenum{3} & 173$\degsym$ & \polenum{4} & 173$\degsym$ & 173$\degsym$ \\
(2144) Marietta      &  (49$\degsym$;$+$87$\degsym$) & \polenum{1} &  19$\degsym$ & \polenum{2} &  19$\degsym$ &  19$\degsym$ \\
(2209) Tianjin       &  (61$\degsym$;$+$87$\degsym$) & \polenum{1} &  20$\degsym$ & \polenum{2} &  19$\degsym$ &  19$\degsym$ \\
\bottomrule
\end{tabular*}
\begin{flushleft}
\noteindex{a} The orbit pole is 
($\Omega - $90$\degsym$;\,90$\degsym - i$)
where
the values of the osculating orbit elements used,
longitude of ascending node $\Omega$
and inclination $i$,
are ecliptic and equinox J2000.0 for
epoch JD~2451800.5
\protect \citep{EMP2000}.
\end{flushleft}
\end{table*}

Table~\ref{AXIAL-RATIOS-TBL}
summarizes very coarse estimates of axial ratios for the CI
model shapes,
noting that the model shapes for the two symmetric poles for each
object
are essentially identical except for being mirror-imaged.
These axial ratios are
very approximate, with uncertainties of at least $\pm 0.1$;
due to the limited number of data and observing geometries,
the pole coordinates are considerably better constrained than are the
derived shape models.
In Table~{\ref{AXIAL-RATIOS-TBL}} for
each object the first pair of model axial ratios given is based on
spatial extents along Cartesian axes having
the $x$- and $z$-directions aligned with
the longest and polar axes, respectively.
The second pair of axial ratios for each model is based on
fitted principal axes of moments of inertia,
assuming uniform density inside the volume enclosed by the
model shape \citep{DRUM14}.
The moments ratios agree with the spatial ratios
to within 0.1, nominally corroborating the estimated uncertainty.
The moments ratios also should be more physically meaningful
than the spatial ratios,
mindful that the moments remain subject to systematic effects
if large-scale non-convex irregularities and/or craters
are present on the asteroid,
because such features will be filled in as flattened regions
on the ``photometric convex hull'' model shape
from convex inversion.
In the final column $\theta_{\rm MI}$ is the angular difference between
the shortest principal axis calculated from the moments
and the polar axis of the model shape.
Caution is needed in interpreting deviations for models
for which $b/c$ is not large enough for the shape modeling to
clearly distinguish the relative lengths of its $b$ and $c$ axes.

\begin{table}
\caption{Model axial ratios for the
spin vector results given in Table~\protect\ref{ALL-SV-RESULTS-TBL}.}
\label{AXIAL-RATIOS-TBL}
\begin{tabular*}{\tblwidth}{@{}llllllr@{}l@{}}
  \toprule
Asteroid             & \multicolumn{2}{c}{Spatial} && \multicolumn{4}{c}{Moments} \\
                     & \multicolumn{2}{c}{extents} && \multicolumn{4}{c}{of inertia} \\ \cline{2-3} \cline{5-8}
                     & $a/b$ & $b/c$ && $a/b$ & $b/c$ & \multicolumn{2}{c}{$\theta_{\rm MI}$} \\
\midrule
 (658) Asteria       & 1.5 & 1.0 && 1.5 & 1.0 & 6&$\degsym$\noteindex{b} \\
 (761) Brendelia     & 1.3 & 1.3 && 1.3 & 1.4 & 2&$\degsym$ \\
 (811) Nauheima      & 1.1 & 1.5 && 1.0 & 1.6 & 2&$\degsym$ \\
 (975) Perseverantia & 1.1 & 1.0 && 1.2 & 1.0 & 8&$\degsym$\noteindex{b} \\
(1029) La~Plata      & 1.4 & 1.4 && 1.5 & 1.3 & 2&$\degsym$ \\ 
(1079) Mimosa        & 1.1 & 1.2 && 1.2 & 1.2 & 9&$\degsym$ \\ 
(1100) Arnica        & 1.2 & 1.0 && 1.2 & 1.0 & 15&$\degsym$\noteindex{b} \\ 
(1245) Calvinia      & 1.5 & 1.0 && 1.6 & 1.0 & 30&$\degsym$\noteindex{b} \\ 
(1336) Zeelandia     & 1.4 & 1.2 && 1.4 & 1.3 & 14&$\degsym$ \\
(1350) Rosselia      & 1.5 & 1.2\noteindex{a} && 1.5 & 1.2\noteindex{a} & 12&$\degsym$ \\ 
(1423) Jose          & 1.7 & 1.0\noteindex{a} && 1.7 & 1.1\noteindex{a} & 55&$\degsym$\noteindex{b} \\ 
(1482) Sebastiana    & 1.6 & 1.1\noteindex{a} && 1.6 & 1.1\noteindex{a} & 6&$\degsym$\noteindex{b} \\
(1618) Dawn          & 1.6 & 1.0 && 1.6 & 1.0 & 45&$\degsym$\noteindex{b} \\ 
(1635) Bohrmann      & 1.4 & 1.3 && 1.3 & 1.3 & 17&$\degsym$ \\
(1725) CrAO          & 1.2 & 1.3 && 1.2 & 1.4 & 2&$\degsym$ \\
(1742) Schaifers     & 2.1 & 1.1 && 2.0 & 1.1 & 23&$\degsym$\noteindex{b} \\
(1848) Delvaux       & 1.7 & 1.4\noteindex{a} && 1.7 & 1.4\noteindex{a} & 2&$\degsym$ \\
(2144) Marietta      & 1.4 & 1.0 && 1.3 & 1.0 & 36&$\degsym$\noteindex{b} \\ 
(2209) Tianjin       & 1.3 & 1.3 && 1.3 & 1.2 & 3&$\degsym$ \\
\bottomrule
\end{tabular*}
\begin{flushleft}
\noteindex{a}
$b/c$ is less reliably determined for this object that has a
high-latitude spin axis.
\newline
\noteindex{b}
Caution is needed in interpreting the deviation
$\theta_{\rm MI}$
for this model that has $b/c \lesssim$ 1.1.
\end{flushleft}
\end{table}

\begin{figure*}
\centering
\includegraphics[width=140mm,height=136mm]{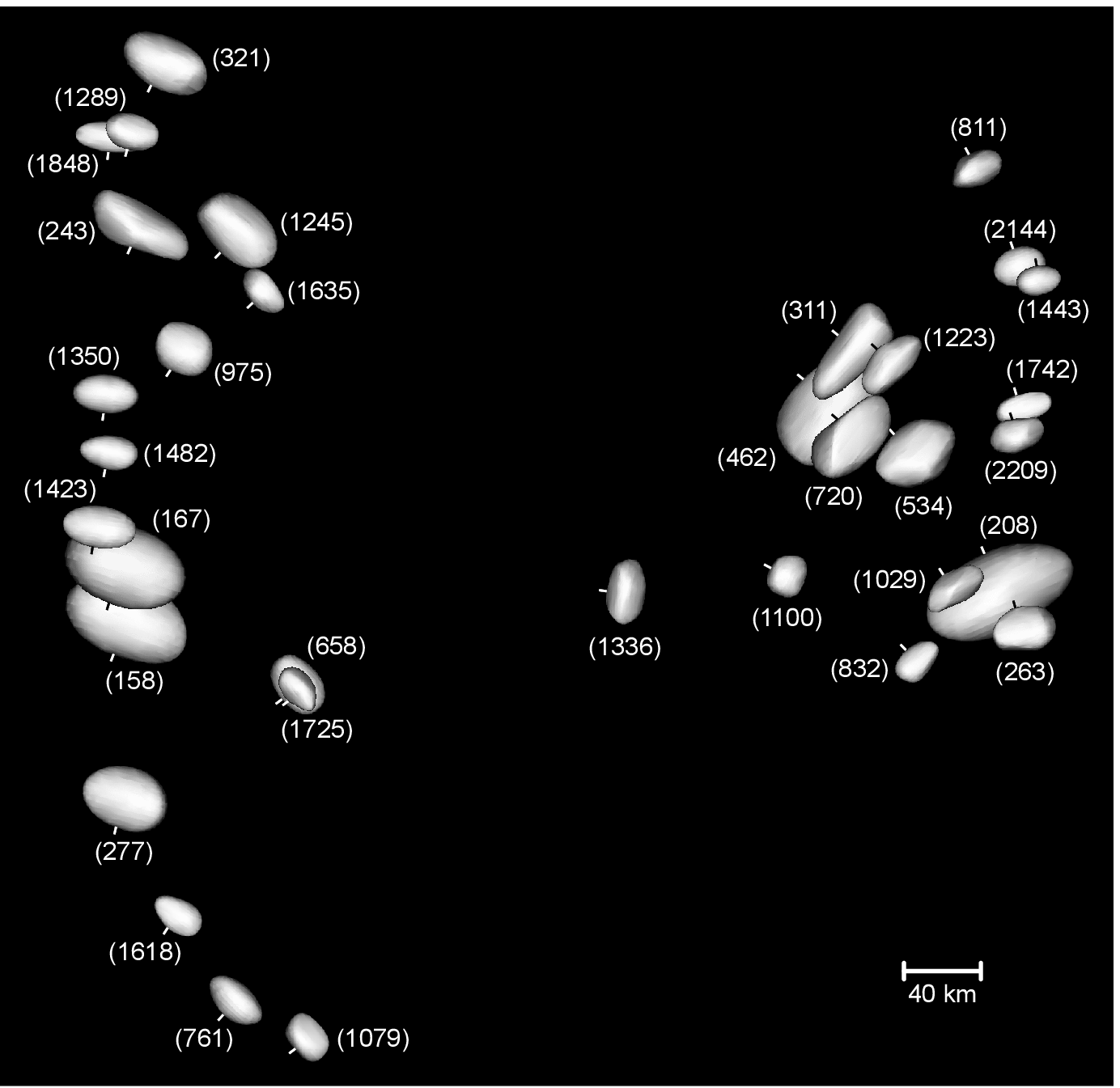}
\caption{Shape model renderings of the 34 objects
comprising the Koronis family spin vector sample,
scaled to represent approximate sizes.
The overall linear scale is set by
the long dimension of (243) Ida \protect\citep{BELT96},
and
each rendering's relative size is calculated using
its model axial ratios so that
the ellipsoidal cross-sectional area of the broadside equatorial view
is proportional to the flux ratio corresponding to
its catalog \filter{H} magnitude difference from Ida.
Views are equatorial and broadside to the longest axis,
fully illuminated using an artificial scattering law
to reveal shape features.
A tick mark on each rendering indicates the object's north pole,
with its position angle equal to the spin obliquity.
The objects are arranged as in
the graph shown in Fig.~\protect\ref{FREQ-VS-OBL-FIG}
so that rotation frequency increases toward the top and
spin obliquity increases numerically toward the left.}
\label{SHAPES-FIG}
\end{figure*}

Renderings of the shape models themselves are
presented
in Fig.~\ref{SHAPES-FIG},
mindful that details of the shape results from these relatively
small lightcurve datasets are not very robust.

The spin vector directions of the four program objects
(1350)~Rosselia, (1423)~Jose, (1482)~Sebastiana, and (1848)~Delvaux
each lie within $15\degsym$ of an ecliptic pole.
As Koronis family members these objects share small orbit
inclinations,
thus their high-latitude spin vectors
restrict Earth-based viewing aspects to be always close
to equatorial.
Because the
lightcurves lack the information needed
to properly scale the model shape in the direction along its polar
axis,
its $b/c$ axial ratio outcome
is sensitive to the shape of the initial ellipsoid.
The reported value of the $b/c$ ratio in each of these cases
was identified 
based on the relative fit $\chi^2$ values
from among a series of trial inversion runs
differing only by the $b/c$ value
supplied for the initial ellipsoid.

The spin vector analyses of
the 19 program objects
are each discussed individually in the following sections.

\clearpage

\subsection{(658) Asteria}

The Asteria lightcurve data set includes six apparitions spanning 24
years,
with a 150-degree gap in aspect ecliptic longitude coverage
centered near $245\degsym$.
Non-overlapping segments of relative photometry from 1999
\citep{SLIV08a} do not permit confidently determining an epoch
for the initial analysis for the sidereal period,
even though the lightcurves combined record about half of a rotation.
Epoch information available from the other five apparitions
nevertheless constrains the sidereal rotation count to three
possibilities,
from which SPA identifies the correct period and retrograde
spin.
The maximum epoch interval corresponds to
9967
sidereal rotations.

\begin{figure}
\centering
\includegraphics[scale=1.00]{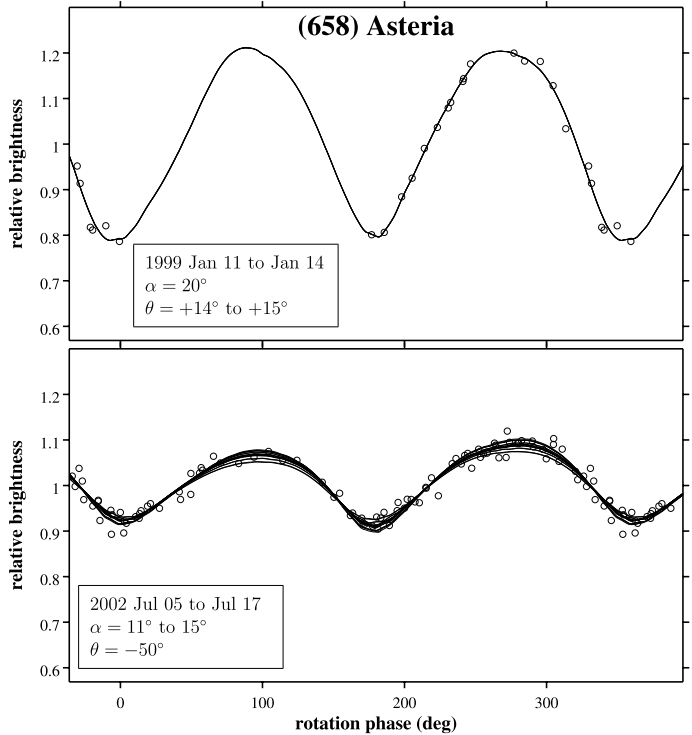}
\caption{Selected lightcurve fits for (658) Asteria
with a pole \polenum{4} at (306$\degsym$;$-$39$\degsym$), plotted as
brightness vs. sidereal rotation phase in degrees.
Data are shown as open circles,
and solid curves represent model brightnesses.
Changes in the underlying lightcurve shape during the course of the observations
appear as non-overlapping model curves.
The legends give the UT dates of the observations
and the corresponding ranges of
solar phase angles $\alpha$
and sub-PAB latitudes $\theta$;
the graphs are presented in order of decreasing $\theta$.
The RMS error of the fit to the combined data set of lightcurves
corresponds to 0.012 mag.}
\label{A0658-P4-CIRESID-FIG}
\end{figure}

Using the lightcurves from all six apparitions,
CI finds a pair of pole solution regions having the expected
symmetry.
Selected lightcurve fits are presented in
Fig.~\ref{A0658-P4-CIRESID-FIG}.

\subsection{(761) Brendelia}
\label{SV-FIRSTNOTES-SEC}

The lightcurves of Brendelia from the 2001 apparition \citep{SLIV08a}
are too incomplete for Fourier series fitting and were not
included in the analysis of epochs for
the sidereal rotation period.
Epochs from 2003--2013 determine an unambiguous sidereal rotation count,
from which SPA identifies retrograde rotation.
The maximum epoch interval corresponds to
1540
sidereal rotations.

\begin{figure}
\centering
\includegraphics[scale=1.00]{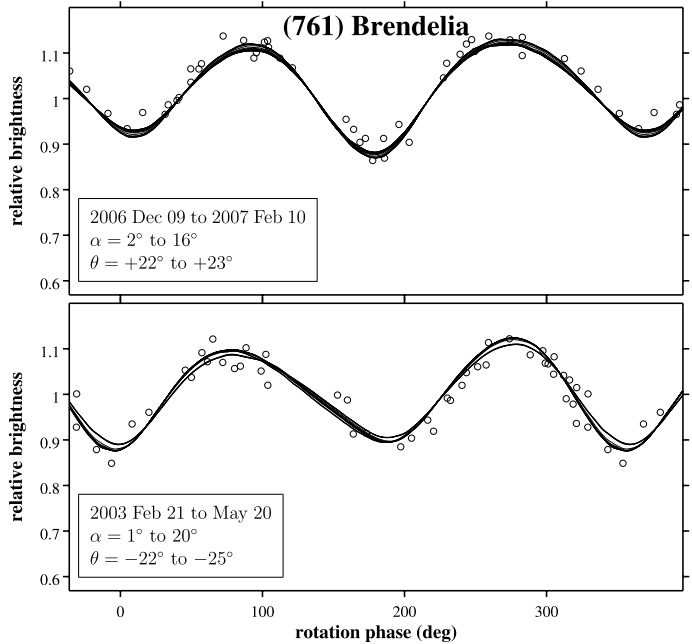}
\caption{The same as Fig.~\protect\ref{A0658-P4-CIRESID-FIG} but for
(761) Brendelia with a pole \polenum{4} at (212$\degsym$;$-$48$\degsym$).
The RMS error of the fit corresponds to 0.026 mag.}
\label{A0761-P4-CIRESID-FIG}
\end{figure}

Brendelia's 58-h rotation period precludes analyzing the
uncombined lightcurves as relative photometry;
instead, the standard-calibrated \filter{R} data
were transformed to the same brightness zero-point as the \filter{V} data
using color information from Table~\ref{CI-RESULTS-TBL}
and analyzed together as standard-calibrated photometry.
Analyzing the calibrated lightcurves
clearly locates the pole regions,
and selected lightcurve fits are presented in
Fig.~\ref{A0761-P4-CIRESID-FIG}.
Our sidereal period for Brendelia agrees with that of
\citet{DURE18a},
and our \polenum{4} is 5 degrees of arc from their
single pole solution.

\subsection{(811) Nauheima}

Lightcurves of Nauheima are available from seven apparitions
spanning 23 years with good coverage in PAB longitude.
The asymmetric shapes and consistently rather low amplitude of the
lightcurves make it harder to accurately locate features that are
related to low-order asteroid shape.
Intervals between the filtered epochs determine an
unambiguous sidereal rotation count,
for which SPA favors the prograde period solution;
the maximum epoch interval corresponds to
49129
sidereal rotations.

The RMS error of the SPA fit to the epochs is larger than is typical
for the corresponding analysis for the other sample objects,
and is dominated by the residual for the epoch
estimated from the incomplete lightcurve from 1989.
Before proceeding with CI the rotation counting was checked by directly
comparing similarly-shaped lightcurves to determine intervals,
an approach that is insensitive to systematic errors that may be
present in the identifications of individual epochs.
The lightcurves from the
apparitions of 2004 and 2005
have similar shapes,
constraining the sidereal period to two candidate ranges within the synodic
constraint of 4.0006 to 4.0016~h.
The interval elapsed between the reflex aspect pair of 2004 and 2007
rules out one of the ranges and reduces the span of the other,
which is then further narrowed first by the similar-aspect pair of
1988 and 1998,
and then finally by the maximum interval of 1984 and 2007
to corroborate the sidereal rotation counts determined from the
Fourier-filtered epochs.

\begin{figure}
\centering
\includegraphics[scale=1.00]{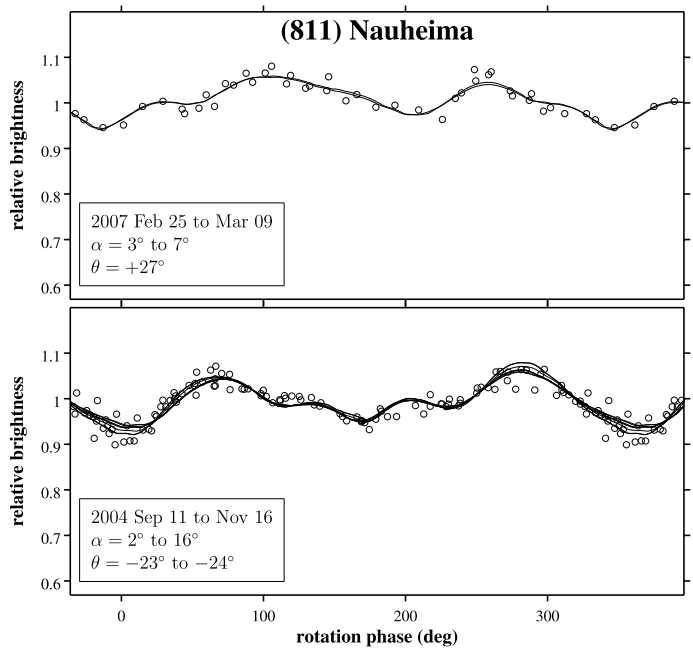}
\caption{The same as Fig.~\protect\ref{A0658-P4-CIRESID-FIG} but for
(811) Nauheima with a pole \polenum{2} at (343$\degsym$;$+$60$\degsym$).
The RMS error of the fit corresponds to 0.018 mag.}
\label{A0811-P2-CIRESID-FIG}
\end{figure}

The observed viewing aspects of Nauheima are clustered near four
ecliptic longitudes,
and weighting the Nauheima lightcurve data on the basis of epochs for
CI input produces a very lopsided distribution in aspect coverage;
to aid the CI analyses for this object,
the weighting was determined based on viewing aspects.
CI finds the pole regions at ecliptic longitudes about $50\degsym$
different from those found by SPA;
the CI locations are favored here because they incorporate lightcurve
amplitude information about pole longitude that is unavailable
to SPA.
The pole-on view of the shape model has a triangular profile,
and selected lightcurve fits are presented in
Fig.~\ref{A0811-P2-CIRESID-FIG}.

\clearpage

\subsection{(975) Perseverantia}

The combined data set of
Perseverantia lightcurves reported in this work
together with those summarized by \citet{SLIV08a}
spans more than 11 years.
\filter{J}-band observations by \citet{VEED95} made on two nights
during one more apparition in 1990 yield a folded composite lightcurve
covering about a quarter of a rotation,
just enough to locate an additional epoch that nearly doubles the span
of observations to almost 23 years.

\begin{figure}
\centering
\includegraphics[scale=1.00]{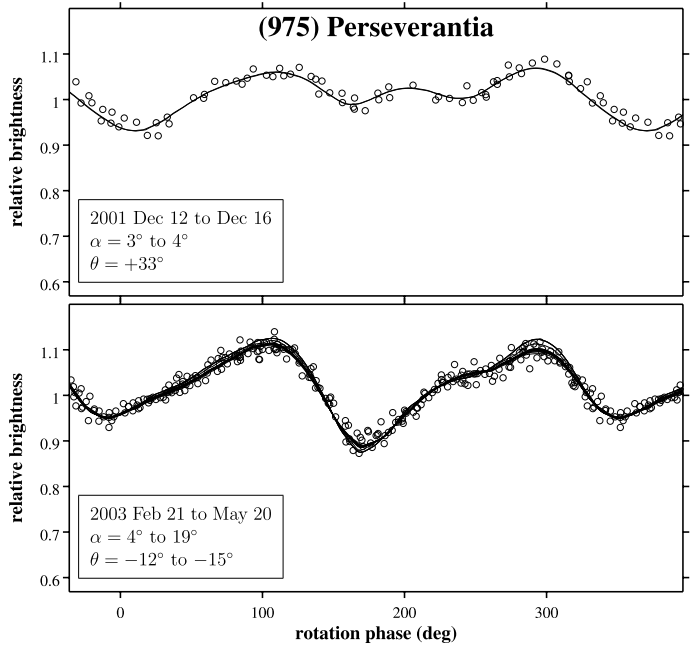}
\caption{The same as Fig.~\protect\ref{A0658-P4-CIRESID-FIG} but for
(975) Perseverantia with a pole \polenum{4} at (238$\degsym$;$-$58$\degsym$).
The RMS error of the fit corresponds to 0.014 mag.}
\label{A0975-P4-CIRESID-FIG}
\end{figure}

SPA analysis of the epochs constrains the sidereal period to
a single pair of candidate solutions with corresponding pole regions,
favoring the retrograde solution over the prograde.
CI confirms that the retrograde poles are correct,
indicating two oval regions exhibiting the expected
symmetry.
The maximum epoch interval corresponds to 27516 sidereal
rotations.
Selected lightcurve fits presented in
Fig.~\ref{A0975-P4-CIRESID-FIG}.

Our sidereal period for Perseverantia agrees with a model reported by
\citet{DURE19},
but our pair of pole solutions is about
$12\degsym$ farther north in latitude than is their pair.

\subsection{(1029) La~Plata}

\begin{figure}
\centering
\includegraphics[scale=1.00]{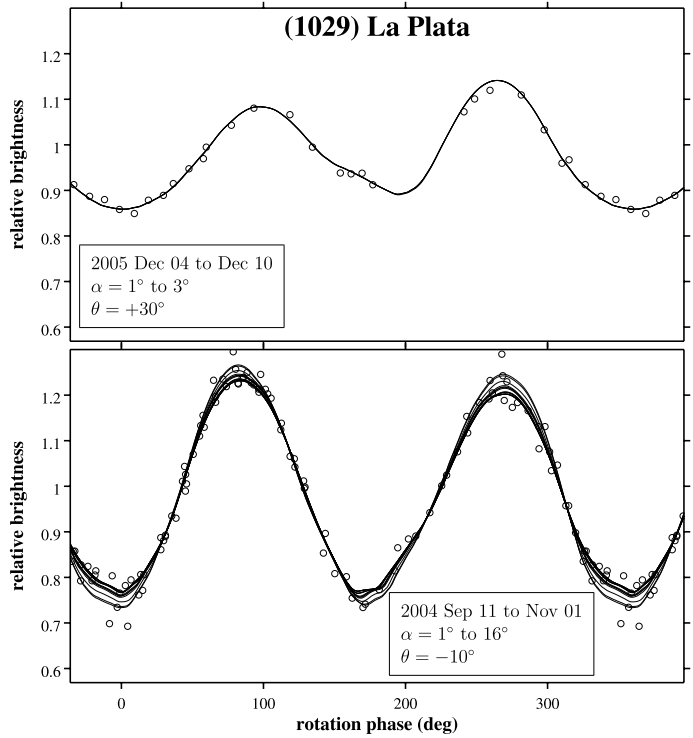}
\caption{The same as Fig.~\protect\ref{A0658-P4-CIRESID-FIG} but for
(1029) La~Plata with a pole \polenum{2} at (281$\degsym$;$+$55$\degsym$).
The RMS error of the fit corresponds to 0.021 mag.}
\label{A1029-P2-CIRESID-FIG}
\end{figure}

Lightcurves of La~Plata span over 31 years,
starting with the 1975 photographic lightcurve from \citet{LAGE78}.
The epochs analyses constrain the sidereal
period to a single prograde solution,
for which CI locates oval pole regions having the expected symmetry.
The maximum epoch interval corresponds to 18007 sidereal
rotations, and
selected lightcurve fits are presented in
Fig.~\ref{A1029-P2-CIRESID-FIG}.

Our results for La~Plata do not agree with those reported by
\citet{DURE18b}---our sidereal period is significantly longer,
and while specific error estimates for their pole do not seem to
have been reported,
our pole locations \polenum{1} and \polenum{2} differ from theirs
by 12 and 33 degrees of arc, respectively.
It is instructive to note that
their period value is close to an alias period in our analysis, and
their pair of pole solutions is not symmetric with
respect to the PGC pole;
instead its 28 degrees of arc departure from symmetry indicates that
at least one of its pole locations is not correct.
\citet{DURE19} revisited the modeling of La Plata and reported
a single pole that is $7\degsym$ from our \polenum{2},
and a sidereal period that is plausibly consistent within
its order of magnitude error estimate.

\clearpage

\subsection{(1079) Mimosa}

The Mimosa lightcurve data set includes eight apparitions spanning
nearly 38 years.
Epoch information estimated from the incomplete 2003 lightcurve
was assigned a larger uncertainty;
low-amplitude lightcurves recorded in 1983 \citep{BINZ87} and in 2017
lack features clearly related to low-order shape and
were not included in epochs analysis.
Epochs from the six apparitions included in the initial analysis
constrain the sidereal rotation count to two possibilities,
one prograde and one retrograde,
differing by half a rotation over the observed interval.
SPA locates a pair of symmetric pole regions for each candidate
rotation count, favoring the retrograde solutions.

\begin{figure}
\centering
\includegraphics[scale=1.00]{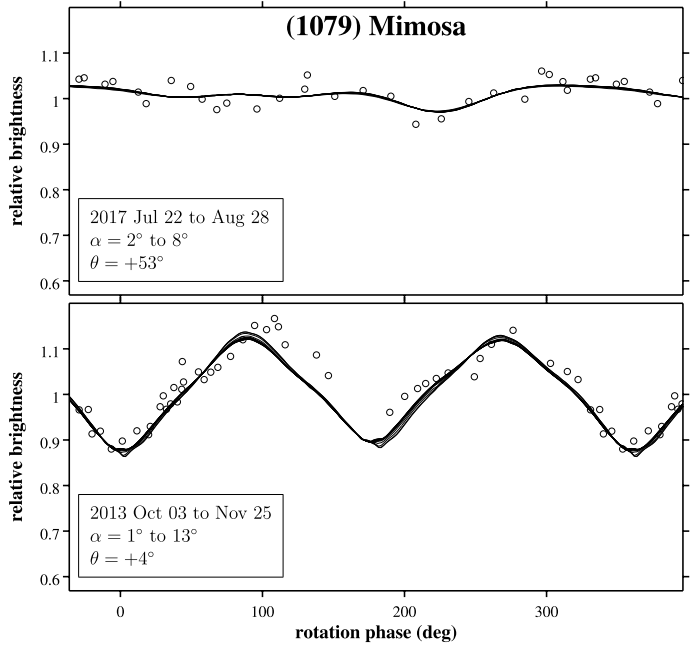}
\caption{The same as Fig.~\protect\ref{A0658-P4-CIRESID-FIG} but for
(1079) Mimosa with a pole \polenum{3} at (137$\degsym$;$-$37$\degsym$).
The RMS error of the fit corresponds to 0.033 mag.}
\label{A1079-P3-CIRESID-FIG}
\end{figure}

Mimosa's 65-h rotation period precludes analyzing the
uncombined lightcurves as relative photometry;
instead, the standard-calibrated \filter{B}, \filter{V}, and \filter{r'} data
were transformed to the same brightness zero-point as the \filter{R} data
and analyzed together as standard-calibrated photometry
using \colorindex{B}{V}  = $0.81  \pm 0.02 $ \citep{BINZ87},
\colorindex{V}{R}  = $0.477 \pm 0.014$ \citep{SLIV08a},
and \colorindex{r'}{R} from Table~\ref{CI-RESULTS-TBL}.
Using the lightcurves from all eight apparitions,
CI locates the two pairs of candidate pole regions corresponding to SPA,
plus two rejected spurious pairs closer to the ecliptic
which would require nonphysical model shapes having $b/c<1$.
CI by itself does not resolve the direction of spin
from among the remaining candidate poles,
likely as a consequence of having only limited information available
at viewing geometries away from the higher-amplitude aspects.
Nevertheless,
the strong correspondence of the CI retrograde regions locations
with the SPA regions,
and the poorer correspondence of the prograde regions,
corroborate the SPA preference for retrograde rotation.
The maximum epoch interval corresponds to 4967 sidereal rotations,
and selected lightcurve fits are presented in
Fig.~\ref{A1079-P3-CIRESID-FIG}.

Our sidereal period for Mimosa is significantly shorter than a result
reported by \citet{DURE20};
their value is close to an alias period in our analysis.
Our pole solutions agree with theirs in latitude,
but are smaller in longitude by 7 to 8 degrees of arc
which is between 2 and 3 times their estimated errors.

\subsection{(1100) Arnica}

Lightcurve data recorded during the nine observed apparitions of
Arnica are clustered near four different viewing aspects.
While the lightcurves from aspect ecliptic longitudes near $25\degsym$
are doubly-periodic and relatively symmetric with amplitudes larger
than 0.2 mag.,
near $185\degsym$ the lightcurves are asymmetric with a distorted
fainter maximum.
At the aspects near $105\degsym$ and $280\degsym$ the lightcurve is
singly periodic with amplitude not exceeding 0.15 mag.

Reliable identification of lightcurve features for epochs
determination is limited to the larger-amplitude apparitions.
The epochs analysis includes unpublished lightcurves recorded during
the 2005 apparition (data provided by R.\ Roy and R.\ Behrend)
to establish a longer maximum time interval between well-defined epochs.
Although applying the Fourier filtering approach to the
singly-periodic lightcurves does yield credible epochs,
they cannot simply be combined in a single analysis
with the epochs from the doubly-periodic aspects,
because epochs from lightcurves having such significantly different shapes
are unlikely to
satisfy the assumption that every epoch in the analysis corresponds
to either the same asterocentric longitude or its reflex.
Instead, the two subsets are analyzed separately.
Epochs from the doubly-periodic lightcurves observed in 2005, 2007,
2010, and 2012 constrain the sidereal period to a single pair of
possible values, one each prograde and retrograde.
The epochs from the singly-periodic lightcurves
observed
in 2004, 2009, and 2014
require integer numbers of rotations between them,
which likewise constrains period candidates to a single pair
prograde and retrograde.
Only the prograde candidate period is consistent with both
subsets.

\begin{figure}
\centering
\includegraphics[scale=1.00]{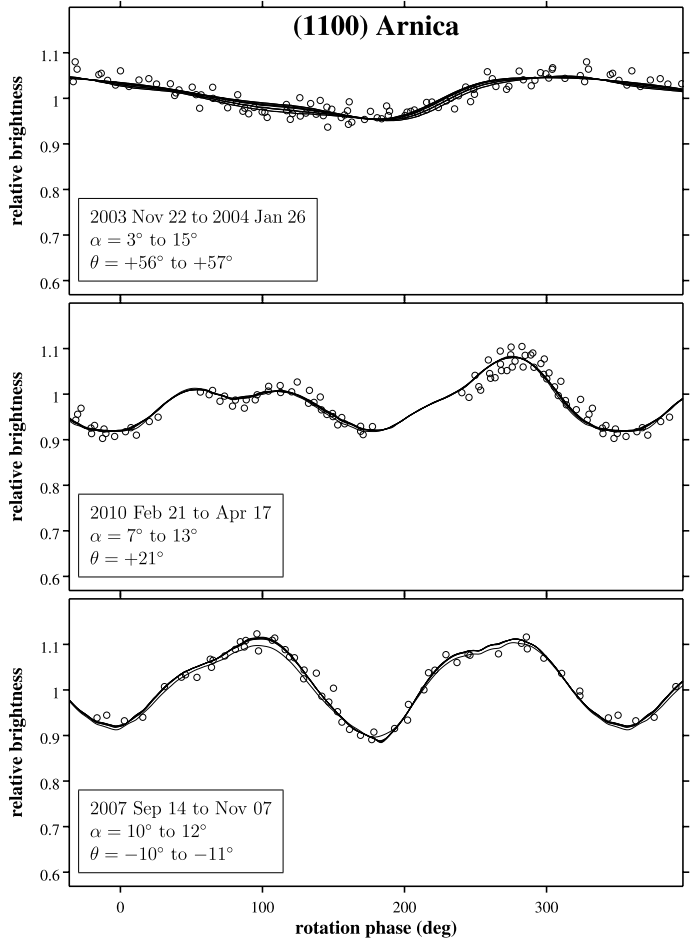}
\caption{The same as Fig.~\protect\ref{A0658-P4-CIRESID-FIG} but for
(1100) Arnica with a pole \polenum{2} at (301$\degsym$;$+$28$\degsym$).
The RMS error of the fit corresponds to 0.016 mag.}
\label{A1100-P2-CIRESID-FIG}
\end{figure}

CI locates the pair of well-defined pole location regions with the
expected symmetry.
To estimate the error in the sidereal period from the combined
analysis of all apparitions,
the RMS error of the SPA fit to the larger-amplitude epochs that span
7.6 years is scaled by the ratio 7.6/15
to account for the larger number of sidereal rotations that occur over
the entire 15-year span of lightcurve data.
The interval between the first and last dates of observations
corresponds to
9079
sidereal rotations,
and
selected lightcurve fits are presented in
Fig.~\ref{A1100-P2-CIRESID-FIG}.

\subsection{(1245) Calvinia}

Before beginning analyses of the Calvinia data,
the inconsistency of the epochs
among the lightcurves from 1977
that was identified by \citet{SLIV08a}
was investigated.
The explanation
is that the photoelectric data of Calvinia from 1977
August that were archived into the Asteroid Photometric Catalogue
\citep{LAGE87} were incorrectly identified there as not being light time
corrected,
when in fact the original data \citep{TEDE79,DEGE79} had already been
light time corrected when first published (E.~Tedesco, personal
communication).
This error has propagated into the current version of the APC
\citep{LAGE11},
in which the time tags themselves for this lightcurve are now
systematically incorrect because there the light time correction has
been applied twice.

The filtered epochs from 2002 through 2012 yield a single possible
count of sidereal rotations,
from which SPA is able to reject the prograde solution leaving an
unambiguous retrograde sidereal period.
Including the data from 1977 further refines the period so
that the maximum epoch interval
corresponds to
63791
sidereal rotations.

\begin{figure}
\centering
\includegraphics[scale=1.00]{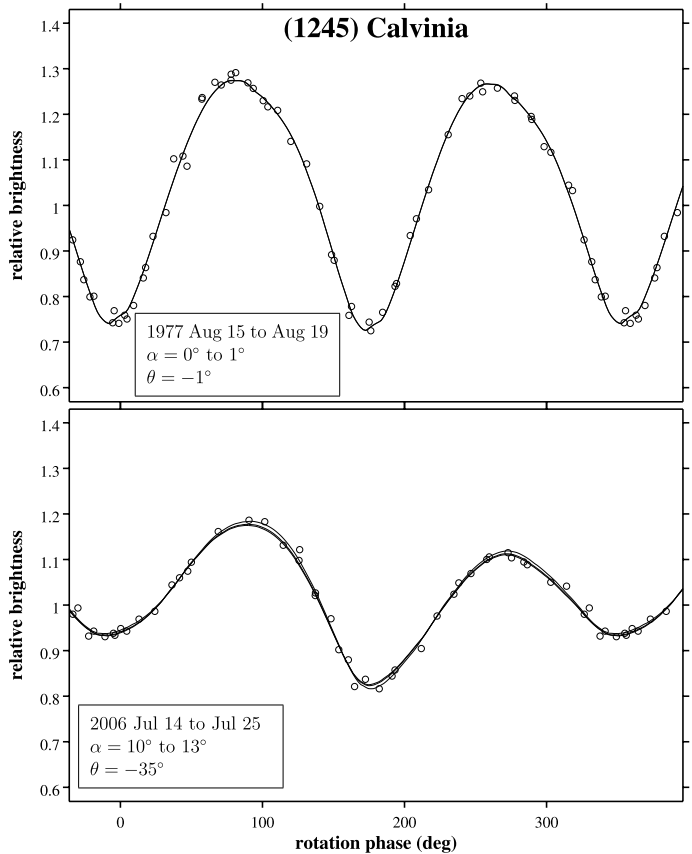}
\caption{The same as Fig.~\protect\ref{A0658-P4-CIRESID-FIG} but for
(1245) Calvinia with a pole \polenum{4} at (235$\degsym$;$-$43$\degsym$).
The RMS error of the fit corresponds to 0.012 mag.}
\label{A1245-P4-CIRESID-FIG}
\end{figure}

CI identifies well-defined pole regions that agree with SPA and
exhibit the expected symmetry.
Selected lightcurve fits are presented in
Fig.~\ref{A1245-P4-CIRESID-FIG}.
The model predicts a lightcurve for the 1975
photographic observations that appears consistent ``by eye'' with the
data,
but those observations were excluded from the analysis because including
them distorts the shape fitting and yields a model that is
inconsistent with the low-noise photoelectric data from 1977.
Our poles and sidereal period for Calvinia
are consistent with
the results reported by \citet{DURE16}.

\subsection{(1336) Zeelandia}

The observed viewing aspects of Zeelandia are broadly clustered near four
ecliptic longitudes,
whose lightcurves exhibit markedly different shapes and amplitudes.
In particular, the lightcurves observed in 2010 are singly-periodic
with low amplitude,
and those from its reflex aspect apparitions in 2007 and 2013
are doubly-periodic but very asymmetric.
Even though those three apparitions initially were excluded
from the analysis of epochs to count sidereal rotations,
epochs from the remaining six apparitions still included were sufficient to
unambiguously determine the rotation count.
Within the constraint of that count result,
the epochs measured from the asymmetric
lightcurves were confirmed to be consistent with predictions
based on the epochs from the symmetric lightcurves
and added to the epochs analyses for SPA,
which indicated pole regions close to the ecliptic
and was able to distinguish the prograde direction of spin.
The maximum epoch interval corresponds to 8545 sidereal rotations.

\begin{figure}
\centering
\includegraphics[scale=1.00]{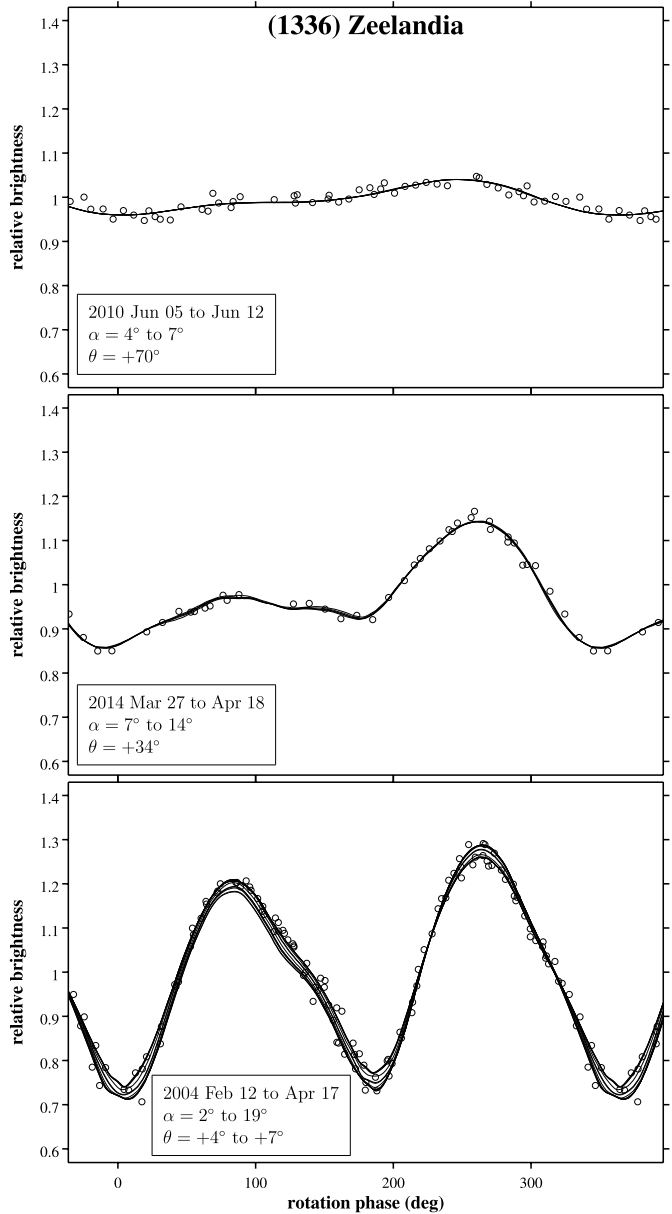}
\caption{The same as Fig.~\protect\ref{A0658-P4-CIRESID-FIG} but for
(1336) Zeelandia with a pole \polenum{1} at (45$\degsym$;$+$5$\degsym$).
The RMS error of the fit corresponds to 0.016 mag.}
\label{A1336-P1-CIRESID-FIG}
\end{figure}

Using CI the two pole regions are very well-defined.
Selected lightcurve fits are presented in
Fig.~\ref{A1336-P1-CIRESID-FIG}.
At near-equatorial aspects the two lightcurve maxima are
comparable in brightness (bottom graph),
but at mid-latitude aspects the maxima are asymmetric,
with one much fainter than the other as it approaches the level of the
minima (center graph).
As is discussed in Sec.~{\ref{BOHR-SEC}} for (1635)~Bohrmann,
this behavior corresponds to the presence of a relatively
large-scale planar region spanning asterocentric mid-latitudes
on the convex model shape.
The 14$\degsym$ deviation of the Zeelandia model's principal axis
from its polar axis suggests that the region
could represent a large concave feature on the asteroid,
such as a crater.

Our sidereal period agrees with
that reported by \citet{DURE19},
however our pole latitudes are
$11\degsym$ farther south and closer to the ecliptic plane.

\subsection{(1350) Rosselia}

Lightcurves of Rosselia span nearly 33 years,
starting with the 1975 photographic lightcurve from \citet{LAGE78}.
The 1991 lightcurve from \citet{VEED95} is too incomplete to locate
extrema and was excluded from the initial sidereal period
determination.
The remaining epochs from 1975 through 2007 identify three candidate
sidereal rotation counts,
from which SPA establishes the correct retrograde solution with pole
regions at high south latitudes and a maximum epoch interval
corresponding to 35280 sidereal rotations.

\begin{figure}
\centering
\includegraphics[scale=1.00]{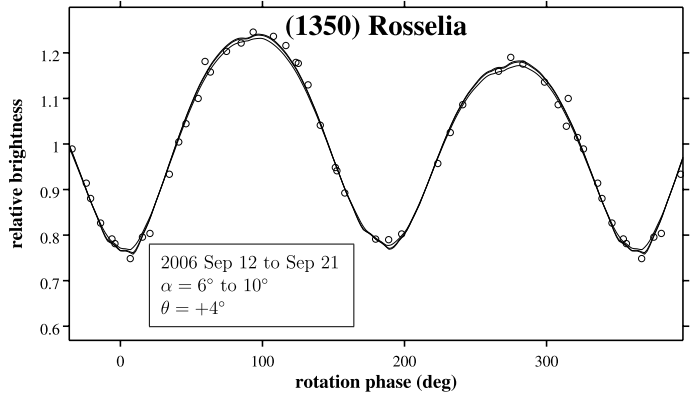}
\caption{The same as Fig.~\protect\ref{A0658-P4-CIRESID-FIG} but for
(1350) Rosselia with a pole \polenum{3} at (131$\degsym$;$-$84$\degsym$).
Only a single graph is selected because
the viewing aspect always is close to equatorial.
The RMS error of the fit corresponds to 0.023 mag.}
\label{A1350-P3-CIRESID-FIG}
\end{figure}

The CI pole regions satisfy the
expected symmetry with respect to the PGC pole,
but their longitudes are only weakly distinguished.
The scant six points comprising the 1991 composite lightcurve were
retained only to confirm consistency of the CI model
and do not affect the results.
Because the pole is at a high latitude,
the model shape $b/c$ axial ratio remains poorly determined
as described in Sec.~\ref{SPINSHAPE-SEC}.
Selected lightcurve fits are presented in
Fig.~\ref{A1350-P3-CIRESID-FIG}.

Pole solutions for Rosselia that have been reported by
\citet{HANU13b} and \citet{DURE16} do not agree with each other,
and our derived pair of spin poles
are about ten
to twenty
degrees farther south in ecliptic latitude
and differ in longitude from all of them.
Our sidereal period agrees with
\citet{HANU13b},
and our \polenum{3} is 13 degrees of arc from
their single pole solution.
\citet{DURE16} report a sidereal period
which is only slightly shorter,
and pole solutions
which are 24 degrees and 23 degrees from our \polenum{3} and
\polenum{4}, respectively.

\subsection{(1423) Jose}

\begin{figure}
\centering
\includegraphics[scale=1.00]{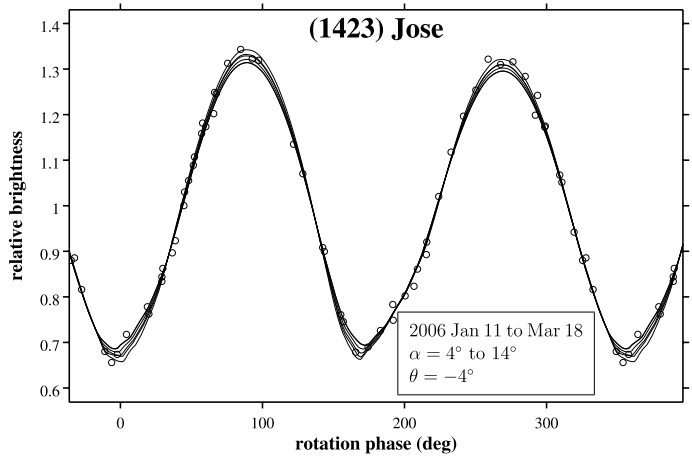}
\caption{The same as Fig.~\protect\ref{A0658-P4-CIRESID-FIG} but for
(1423) Jose with a pole \polenum{3} at (98$\degsym$;$-$80$\degsym$).
Only a single graph is selected because
the viewing aspect always is close to equatorial.
The RMS error of the fit corresponds to 0.024 mag.}
\label{A1423-P3-CIRESID-FIG}
\end{figure}

Lightcurves of Jose are available from five consecutive apparitions,
comprising the only object data set in the analysis sample having data
from fewer than six apparitions.
The epochs are nevertheless sufficient to unambiguously count
sidereal rotations,
and for SPA to identify retrograde spin.
The maximum epoch interval corresponds to
3610
sidereal rotations.
The spin vector solution regions are strongly constrained in
latitude near the south ecliptic pole;
the longitudes are less strongly distinguished but do satisfy the
expected symmetry.
The model shape $b/c$ axial ratio remains poorly determined given the
high latitude of the pole,
as described in Sec.~\ref{SPINSHAPE-SEC}.
Selected lightcurve fits are presented in
Fig.~\ref{A1423-P3-CIRESID-FIG}.
\citet{HANU13b} studied Jose;
our sidereal period agrees with theirs
and our \polenum{3} is 4 degrees of arc from 
their single reported pole solution.

\clearpage

\subsection{(1482) Sebastiana}

\begin{figure}
\centering
\includegraphics[scale=1.00]{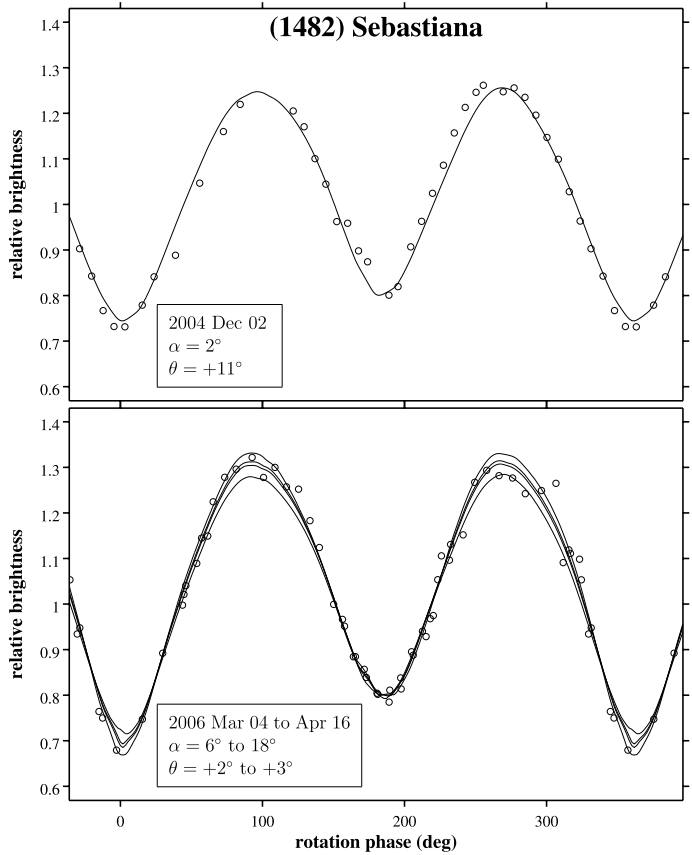}
\caption{The same as Fig.~\protect\ref{A0658-P4-CIRESID-FIG} but for
(1482) Sebastiana with a pole \polenum{4} at (237$\degsym$;$-$79$\degsym$).
The RMS error of the fit corresponds to 0.024 mag.}
\label{A1482-P4-CIRESID-FIG}
\end{figure}

Epochs analysis for Sebastiana determines an unambiguous
sidereal rotation count,
whose retrograde solution is favored by SPA.
The maximum epoch interval corresponds to
18890
sidereal rotations.
Sebastiana was the only program object
for which
the CI convexity regularization weight parameter
needed to be changed from its default value.
Although the shapes of the pole solution regions
are somewhat weakly defined,
the poles are unambiguously near the south ecliptic pole,
leaving the model shape $b/c$ axial ratio poorly determined
as described in Sec.~\ref{SPINSHAPE-SEC}.
Selected lightcurve fits are presented in
Fig.~\ref{A1482-P4-CIRESID-FIG}.

Our sidereal period agrees with the result of
\citet{HANU11} within its reported precision.
Our \polenum{3} and \polenum{4} are 11 and 13 degrees of arc
respectively from their corresponding poles,
differences which are comparable to their own estimated errors.

\subsection{(1618) Dawn}

The Dawn lightcurve data set includes six apparitions spanning
nearly 18 years.
Lightcurves from the 2020 apparition
are too incomplete for Fourier series fitting and were not
included in the analysis of epochs for
the sidereal rotation period.
Epochs from the remaining five apparitions
determine an unambiguous sidereal rotation count,
from which SPA favors retrograde rotation.
The maximum epoch interval corresponds to 3559 sidereal rotations.

Dawn's 43-h rotation period precludes analyzing the uncombined
lightcurves as relative photometry;
instead, the standard-calibrated \filter{V} data were transformed to
the same brightness zero-point as the \filter{R} data
using \colorindex{V}{R} = $0.436 \pm 0.014$ \citep{SLIV08a}
and analyzed together as standard-calibrated photometry.
The data from the apparitions in 2017, 2020, and 2021 that had not
been put onto a standard system were prepared for analysis as
composite lightcurves of relative photometry.

\begin{figure}
\centering
\includegraphics[scale=1.00]{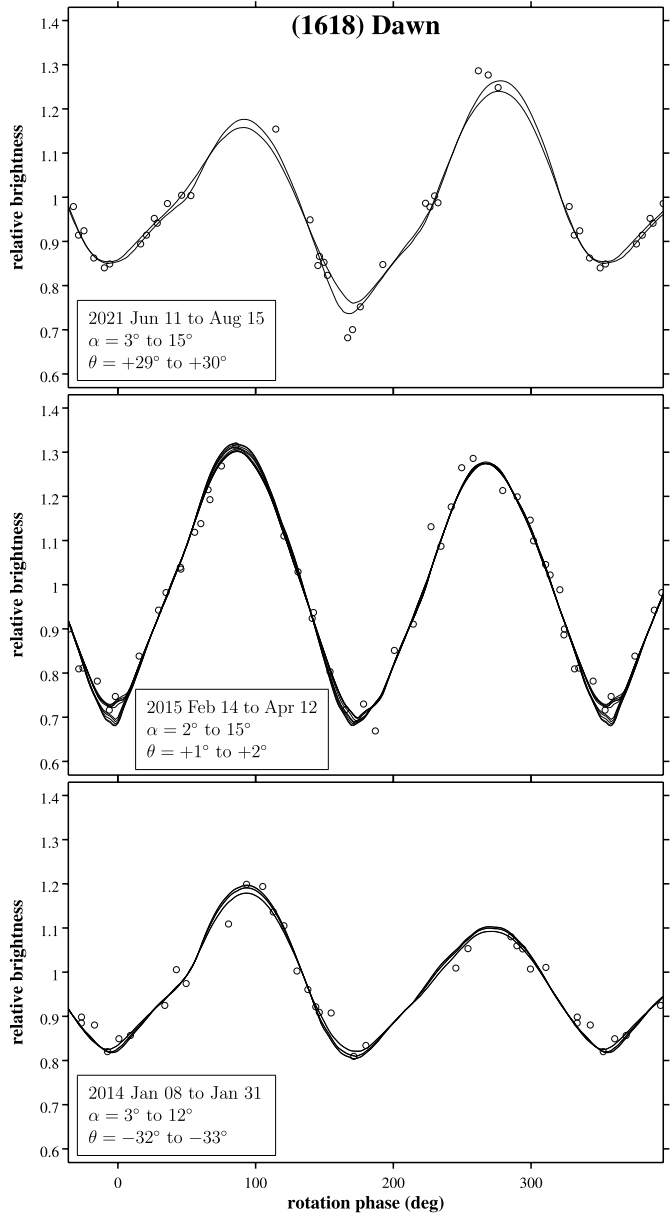}
\caption{The same as Fig.~\protect\ref{A0658-P4-CIRESID-FIG} but for
(1618) Dawn with a pole \polenum{3} at (101$\degsym$;$-$57$\degsym$).
The RMS error of the fit corresponds to 0.027 mag.}
\label{A1618-P3-CIRESID-FIG}
\end{figure}

CI agrees with SPA to favor retrograde rotation,
and locates a pair of retrograde pole solution regions which exhibit
the expected symmetry.
Selected lightcurve fits are presented in
Fig.~\ref{A1618-P3-CIRESID-FIG}.
Our sidereal period for Dawn is consistent with that
reported by \citet{HANU13b} within its error,
and our pair of pole solutions
are about 30 degrees of arc from theirs
which is consistent with their reported uncertainties
of 10 to 20 degrees.

\subsection{(1635) Bohrmann}
\label{BOHR-SEC}

The lightcurves of Bohrmann exhibit markedly different shapes
depending on viewing aspect.
During the similar-aspect apparitions in 2003 and 2008,
and at approximately the reflex observing direction in 2011,
the lightcurves were singly periodic with
one relatively undistorted maximum
comprising half of the rotation phase,
and the other half nearly flat at the minimum
brightness.
To use Fourier series model fits to identify epochs
from these lightcurves despite their almost entirely ``missing''
one of the maxima,
only the half of rotation phase
centered on the undistorted maximum was retained
and fit at half the rotation period.

\begin{figure}
\centering
\includegraphics[scale=1.00]{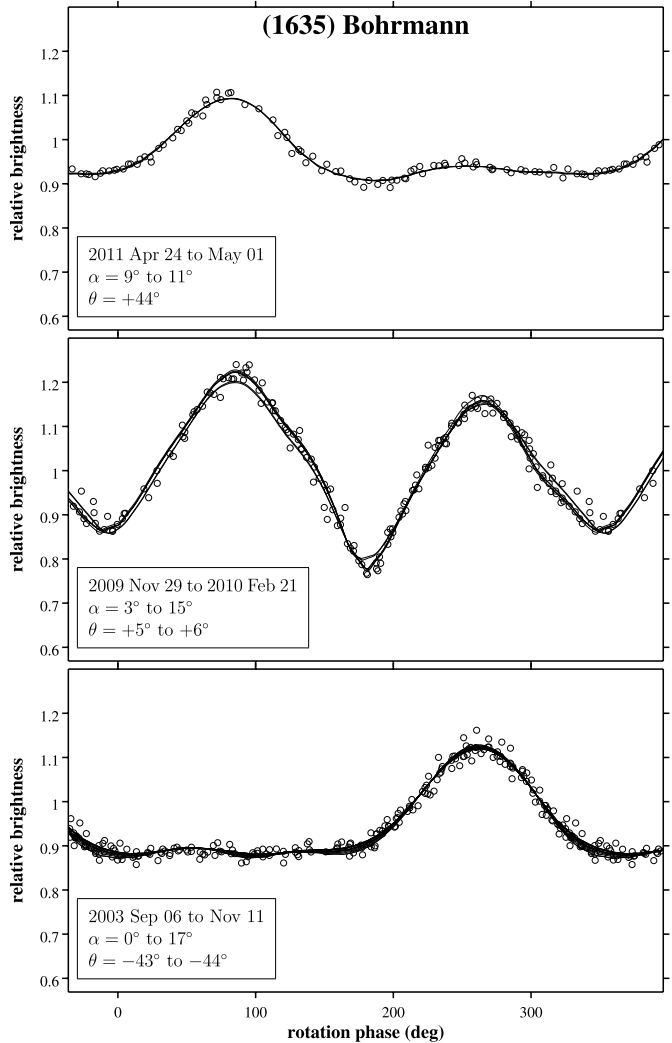}
\caption{The same as Fig.~\protect\ref{A0658-P4-CIRESID-FIG} but for
(1635) Bohrmann with a pole \polenum{3} at (9$\degsym$;$-$46$\degsym$).
The RMS error of the fit corresponds to 0.018 mag.}
\label{A1635-P3-CIRESID-FIG}
\end{figure}

\begin{figure}
\centering
\includegraphics[scale=1.00]{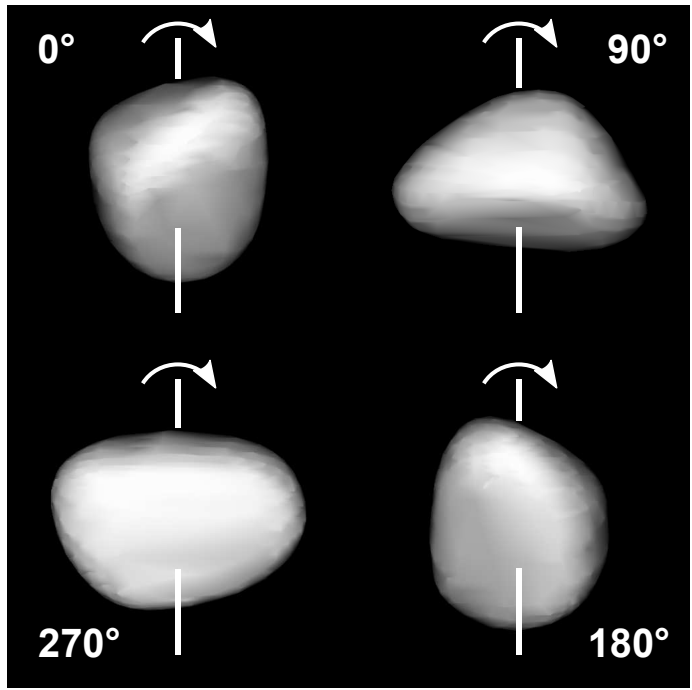}
\caption{Shape model for the {\polenum{3}} solution
of (1635) Bohrmann showing its axis and direction of spin,
rendered for a sub-observer latitude of $-$43$\degsym$
to correspond to the lightcurves in the bottom graph in
Fig.~{\protect\ref{A1635-P3-CIRESID-FIG}}.
At rotation phase 270$\degsym$ the planar region is about 25
degrees from face-on and does not reduce the projected area of the
model for the undistorted lightcurve maximum,
but at what would be the opposite maximum at 90$\degsym$ the
region is foreshortened at about 65 degrees from face-on,
reducing the overall projected area for a lightcurve brightness that
is not much different from the nominal minima at 0$\degsym$ and
180$\degsym$.}
\label{A1635-P3-RENDERINGS-FIG}
\end{figure}

The available epoch data unambiguously count sidereal rotations,
and SPA identifies retrograde
spin.
The maximum epoch interval corresponds to
13219
sidereal rotations.
CI locates well-defined pole regions that are consistent with those
from SPA.
Selected lightcurve fits are presented in
Fig.~\ref{A1635-P3-CIRESID-FIG},
which shows that opposite lightcurve maxima were distorted in
2003 and 2011.
The model shapes show a large-scale planar feature,
and the 17-degree difference between the model's shortest principal axis
and its polar axis is the largest
deviation among those models for which $b/c>1.1$,
suggesting that a concave region on the asteroid,
such as a large crater
corresponding to the planar feature on the model,
is responsible for the distorted lightcurve maxima
as shown in Fig.~{\ref{A1635-P3-RENDERINGS-FIG}}.

Our sidereal period and poles for Bohrmann agree with the
published results of \citet{HANU11}.
Our \polenum{3} and \polenum{4} are respectively
9 and 12 degrees of arc from their poles,
which carry their own estimated errors of about 10 degrees of arc.

\subsection{(1725) CrAO}

Lightcurves of CrAO are available from
seven apparitions spanning 17 years.
The larger-amplitude lightcurves are asymmetric in time
with distorted maxima,
requiring extra care to locate the extrema for epochs analyses.
The smaller-amplitude lightcurve from the 2018 apparition is too
incomplete to confidently locate extrema,
and was excluded from epochs analysis.
Epochs measured from the other six apparitions are sufficient to
determine the sidereal period and identify retrograde spin;
the maximum interval corresponds to
6676
sidereal rotations.

\begin{figure}
\centering
\includegraphics[scale=1.00]{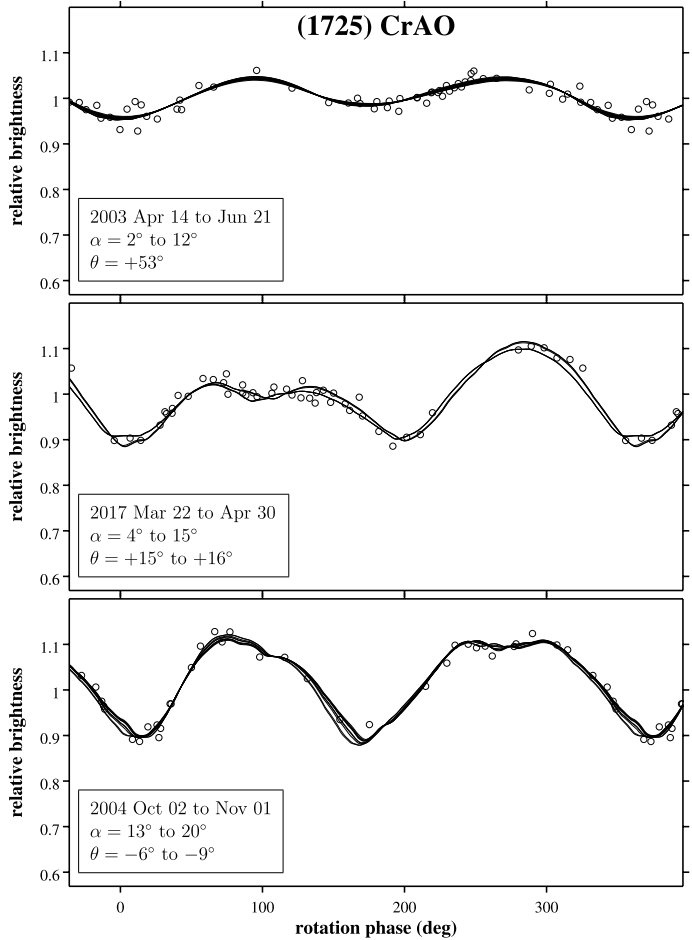}
\caption{The same as Fig.~\protect\ref{A0658-P4-CIRESID-FIG} but for
(1725) CrAO with a pole \polenum{3} at (64$\degsym$;$-$39$\degsym$).
The RMS error of the fit corresponds to 0.013 mag.}
\label{A1725-P3-CIRESID-FIG}
\end{figure}

To include the data from 2018 in the CI analyses,
its short non-overlapping individual lightcurves were combined into a
single lightcurve of relative photometry by referencing all five
nights' measurements to the same comparison star,
and then compositing them in time and in brightness.
Using the lightcurves from all seven apparitions,
CI locates pole regions
that are consistent with those from SPA.
Selected lightcurve fits are presented in
Fig.~\ref{A1725-P3-CIRESID-FIG}.

\subsection{(1742) Schaifers}

Prior to comprehensive analysis of the Schaifers data set,
folding the rather sparsely-sampled lightcurves from 1983
\citep{BINZ87} at the improved synodic rotation period \citep{SLIV08a}
raised doubt about the reality of the unusually large amplitude of the
original published composite lightcurve.
Its asymmetric very deep minimum was recorded only on one night,
whose lightcurve brightness zero-point relative to the composite was
based on a single data point that overlaps other rotation phase
coverage.
Folding at the improved period revealed phase overlap of a second
measurement,
but its brightness zero-point to match the composite is more than 0.3
mag. different from that of the first point,
indicating that at least one of the two brightness measurements is
incorrect.
Shifting the Oct.~9 lightcurve in brightness to agree at the second
phase overlap instead of the first yields a composite lightcurve
with overall shape and smaller amplitude both comparable to those
observed for Schaifers during the other apparitions,
thus we conclude that the brightness of the first overlapping point
likely is incorrect,
and that the lightcurve amplitude at the 1983 observing aspect was not
so extreme.
We note also that there are corroborating hints of possible systematic
errors in the standard calibrations of lightcurves of other objects
that were recorded during the same observing run with Schaifers in
1983
\citep{SLIV09}.

\begin{figure}
\centering
\includegraphics[scale=1.00]{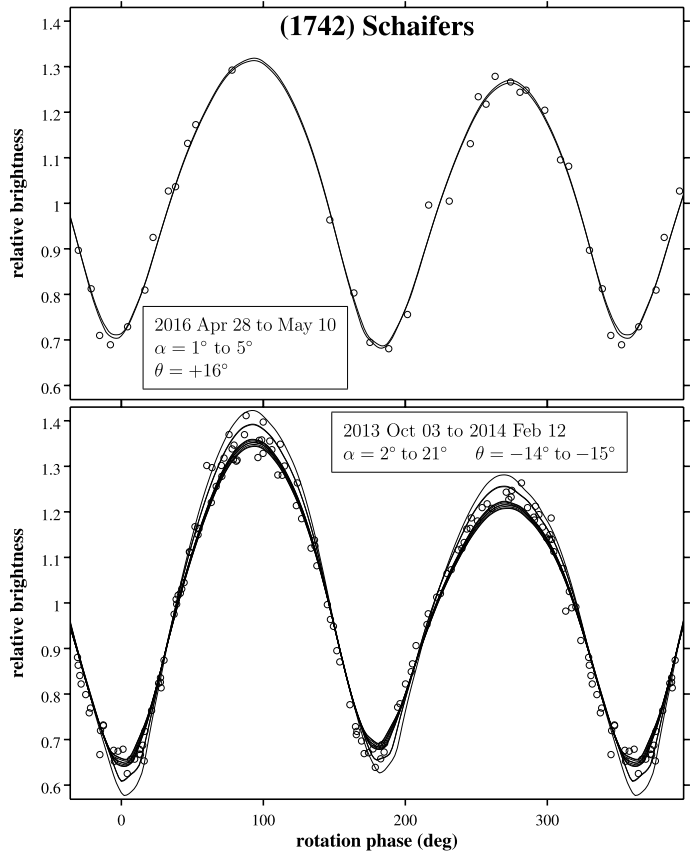}
\caption{The same as Fig.~\protect\ref{A0658-P4-CIRESID-FIG} but for
(1742) Schaifers with a pole \polenum{1} at (27$\degsym$;$+$71$\degsym$).
The RMS error of the fit corresponds to 0.030 mag.}
\label{A1742-P1-CIRESID-FIG}
\end{figure}

The full set of Schaifers epochs constrains the sidereal period to
a single pair of candidate solutions,
favoring the prograde solution over the retrograde.
CI indicates two roughly oval regions which exhibit the expected
symmetry.
The maximum epoch interval corresponds to
33454
sidereal rotations.
Selected lightcurve fits are presented in
Fig.~\ref{A1742-P1-CIRESID-FIG}.

Our sidereal period for Schaifers agrees with the results of
\citet{HANU11} and \citet{HANU13b} to within their published
precisions,
but our poles differ considerably from theirs
by 17 to 23 degrees of arc overall,
individually different
in longitude, in latitude, or in both.
Neither of their pole pairs are consistent with
the expected symmetry with respect to the PGC pole;
instead they depart significantly by 11 to 14 degrees of arc
which suggests that some systematic problem likely is present.

\clearpage

\subsection{(1848) Delvaux}

For spin vector and shape analysis of Delvaux an
unpublished lightcurve recorded during the 2011 apparition
(data provided by R.\ Roy and R.\ Behrend)
was also included.
The lightcurve amplitude during each of the seven observed
apparitions was greater than 0.5 mag,
suggesting a
high-latitude
spin vector and relatively elongated shape.
Epochs analyses yield an unambiguous sidereal period and retrograde
spin,
for which the maximum epoch interval corresponds to 27271 sidereal
rotations.

\begin{figure}
\centering
\includegraphics[scale=1.00]{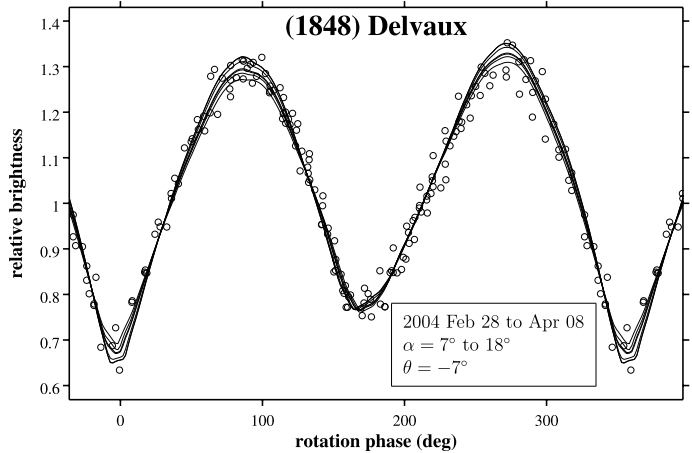}
\caption{The same as Fig.~\protect\ref{A0658-P4-CIRESID-FIG} but for
(1848) Delvaux with a pole \polenum{3} at (143$\degsym$;$-$83$\degsym$).
Only a single graph is selected because
the viewing aspect always is close to equatorial.
The RMS error of the fit corresponds to 0.031 mag.}
\label{A1848-P3-CIRESID-FIG}
\end{figure}

The spin vector is strongly constrained to be near the
south ecliptic pole,
thus the solution region ecliptic longitudes are only weakly
indicated,
and
the model shape $b/c$ axial ratio remains poorly determined
as described in Sec.~\ref{SPINSHAPE-SEC}.
Selected lightcurve fits are presented in
Fig.~\ref{A1848-P3-CIRESID-FIG}.

Our sidereal period for Delvaux agrees with the partial model of
\citet{DURE18a},
but our pole latitudes are more than 15 degrees farther south.
\citet{DURE19}
revisited the modeling of Delvaux and reported
a sidereal period that agrees with ours
but with a pair of poles
having latitudes comparable to the
previous partial model,
$20\degsym$ to $25\degsym$ farther north
than our solutions.

\subsection{(2144) Marietta}

For spin vector and shape analysis of Marietta,
the lightcurves reported in this work
and the data reported by \citet{SLIV08a} and
\citet{ARRE14}
were supplemented
with an unpublished lightcurve recorded during the 2010 apparition
(data provided by R.\ Roy and R.\ Behrend).

The available epoch data yield an unambiguous sidereal rotation
period and prograde spin,
with the maximum epoch interval corresponding to
28074 sidereal rotations.
CI and SPA agree that the pole is at a high north ecliptic latitude,
but the CI pole regions are indistinct.
Fitting from trial initial poles chosen within the regions ``by
eye'' yields a variety of statistically indistinguishable solutions
without consistent convergence to indicate a favored pole location,
and often having peculiar model shapes.

While exploring how
to more deliberately identify an initial value for the pole location
fitting,
it was noticed that the $\chi^2$ contours results from
an analysis of the lightcurve data set using
the simultaneous amplitude-magnitude-aspect (SAM) method of \citet{DRUM88}
resembled the outlines of the CI pole region.
Ultimately an approach described by \citet{SLIV03} which combines the
SPA method with the SAM method was used to choose the initial location
for the pole,
from which the CI fitting converged within one degree of arc.
The pole errors were estimated based on the extents of the portions of
the CI solution regions that satisfy the expected symmetry.

\begin{figure}
\centering
\includegraphics[scale=1.00]{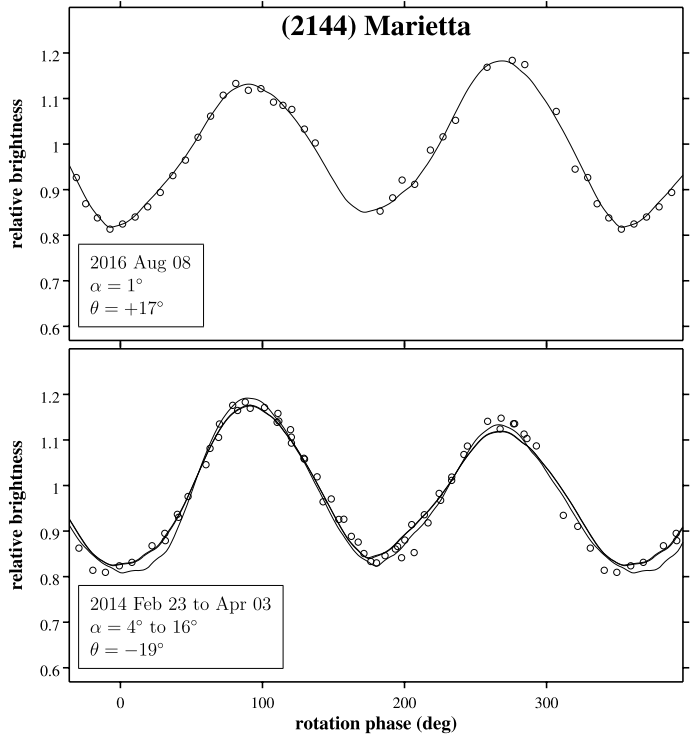}
\caption{The same as Fig.~\protect\ref{A0658-P4-CIRESID-FIG} but for
(2144) Marietta with a pole \polenum{1} at (145$\degsym$;$+$72$\degsym$).
The RMS error of the fit corresponds to 0.021 mag.}
\label{A2144-P1-CIRESID-FIG}
\end{figure}

The converged pole solution for Marietta is at a latitude high enough
that the lightcurves provide only limited information to scale the
model shape in the direction along the polar axis,
leaving the CI shape determination sensitive to the $b/c$ axial ratio
of the initial ellipsoid.
Therefore the same approach that was used to obtain model shapes for the
program objects having the highest-latitude poles also was used for Marietta.
Selected lightcurve fits are presented in
Fig.~\ref{A2144-P1-CIRESID-FIG}.

Our sidereal period agrees with a model solution reported
by \citet{DURE20},
but our pole locations differ from theirs in a curious way:
although the differences are both mainly in ecliptic latitude,
our \polenum{1} 
is south $6\degsym$ ($\sim 3\times$ estimated error)
while
our \polenum{2} 
is in the opposite direction,
north $5\degsym$ ($\sim 2\times$ estimated error).
It is instructive to note that
their pole solutions depart from symmetry with respect to the PGC pole
by 10 degrees of arc,
which suggests that there might be a systematic problem with 
at least one of them.

\subsection{(2209) Tianjin}
\label{SV-LASTNOTES-SEC}

For spin vector and shape analysis of Tianjin the
lightcurves reported in this work
were supplemented with the lightcurves recorded
in 1996 by \citet{FLOR97}.
Analysis of the combined set of epochs unambiguously
identifies the sidereal period and prograde rotation,
for which the maximum epoch interval corresponds to
18703 sidereal rotations.

\begin{figure}
\centering
\includegraphics[scale=1.00]{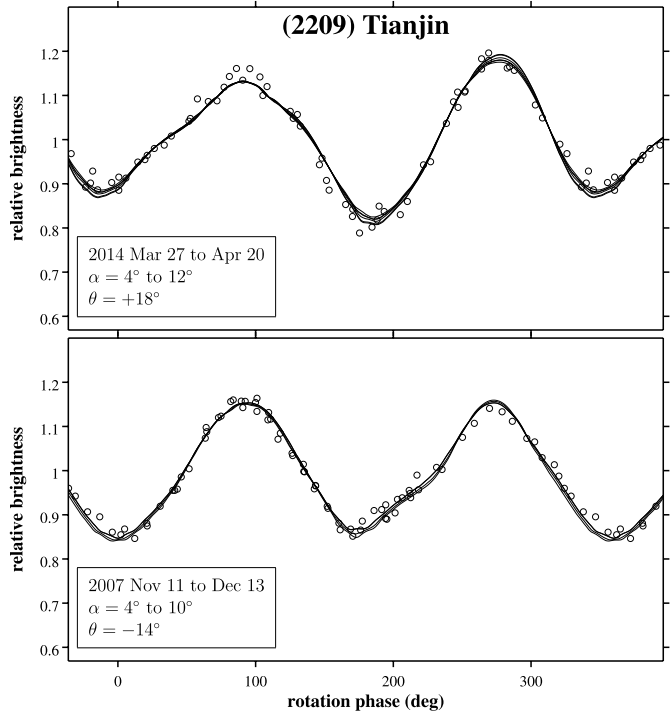}
\caption{The same as Fig.~\protect\ref{A0658-P4-CIRESID-FIG} but for
(2209) Tianjin with a pole \polenum{1} at (19$\degsym$;$+$68$\degsym$).
The RMS error of the fit corresponds to 0.022 mag.}
\label{A2209-P2-CIRESID-FIG}
\end{figure}

Tianjin's spin pole orientation is close enough to the north ecliptic
pole that
the pole longitudes are only weakly distinguished,
and determining the reported $b/c$ axial ratio
of the model shape required the same approach
described in
Sec.~\ref{SPINSHAPE-SEC}
that was used
for the program objects having the highest-latitude poles.
Selected lightcurve fits are presented in
Fig.~\ref{A2209-P2-CIRESID-FIG}.

Our sidereal period agrees with a model solution reported by
\citet{DURE19},
but our \polenum{2} is $12\degsym$ farther north in latitude
and 16 degrees of arc away overall
from their single reported pole location.

\section{Discussion}
\label{DISC-SEC}

\subsection{Comparison with sparse-data period and pole
determinations}

As noted in the Introduction,
spin vector determinations
based on sparse-data analysis approaches
have been reported in
the literature
for fifteen of the observing program objects
\citep{HANU11,HANU13b,DURE16,DURE18a,DURE18b,DURE19,DURE20},
providing an opportunity to compare independent solutions based on the
same underlying modeling assumptions and convex inversion analysis
algorithm,
but using the different data sets.
The sparse-data results were compared individually with their
counterpart results from this work in the descriptions of the program
objects' spin vector analyses in
Secs.~\ref{SV-FIRSTNOTES-SEC}--\ref{SV-LASTNOTES-SEC};
here in this section the comparisons are discussed together as a
population sample.

{\em Sidereal rotation periods.}
Long time intervals available in the
lightcurve data sets assembled for the
observing program objects
increase the numbers of elapsed rotations
and reduce the period uncertainties---four
of our program solution periods are 10 or more times more precise
than the corresponding sparse-data periods,
and fourteen more are between 3 and 10 times improved.
The period results themselves for each object are in mainly good
agreement; 17 of the 19 periods agree within their estimated errors.
The two remaining pairs of period solutions each systematically
differ by five to six times their errors,
corresponding to an 11\% fraction of false positives
per the terminology of \citet{DURE18a}.
Both of these sparse-data periods are close to alias periods
found during the analyses reported in this work.

\begin{figure}
\centering
\includegraphics[scale=1.00]{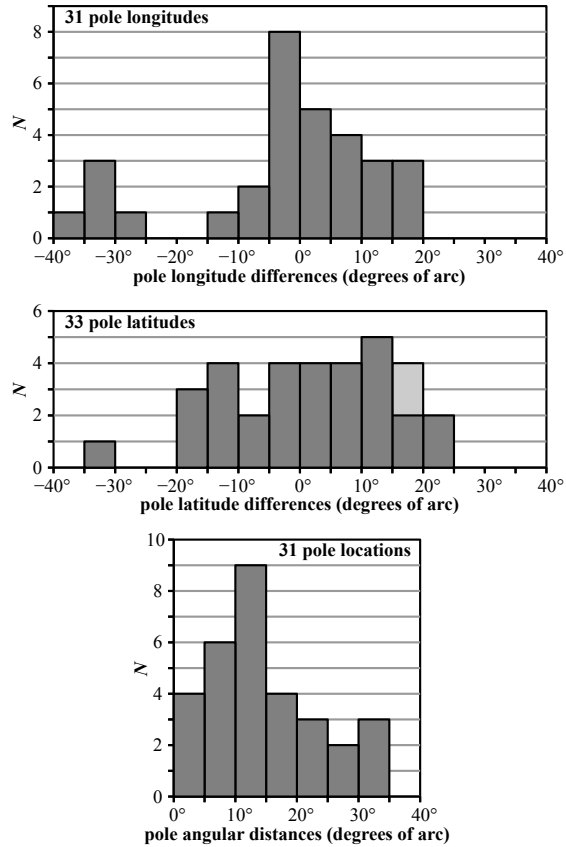}
\caption{Distributions of angular distances between spin poles
from the observing program
and sparse-data results in the literature for the same objects.
(upper graph)
Pole longitude signed differences;
the distribution mean is $-$2 $\pm$ 15 degrees of arc.
(center graph)
Pole latitude signed differences;
lighter grey color
represents differences from sparse-data partial solutions.
The distribution mean is $+$2$\degsym \pm$ 13$\degsym$.
(lower graph)
Unsigned angular distances between pole locations;
the root-mean-squared angular difference is 17 degrees of arc.}
\label{POLEDIFFS-DISTRIBS-FIG}
\end{figure}

{\em Spin pole locations.}
Histograms showing the
distributions of the angular distances between
the spin poles from our observing program
and those based on analyses of sparse data sets
are presented in
Fig.~\ref{POLEDIFFS-DISTRIBS-FIG}.
The small mean signed differences of the pole longitudes
and latitudes of
$-2$ and
$+2$ degrees of arc,
respectively,
indicate that
the agreement of pole coordinates is good in the average.
The differences' RMS dispersions of
$\pm 15$ and
$\pm 13$ degrees of arc,
respectively,
are comparable to each other
and to the RMS dispersion
$\pm 17$ degrees of arc
of the
unsigned angular distances between corresponding pole locations.
The similarity of the dispersions
suggests $\sim 15$ degrees of arc
as an estimated external error for each pole coordinate,
based on the comparison with the pole determinations
from our observing program data sets containing mainly
dense lightcurves.

Comparing the external error
with estimated internal pole errors is not entirely straightforward,
partly because the CI calculations do not admit
analytic formal Gaussian internal errors
for individual pole solutions.
Instead of error estimates for each pole location,
\citet{HANU11} and \citet{HANU13b} report pole errors
as ranges of the typical uncertainties that one can expect from CI
for models based mainly on sparse data.
The estimates in the latter paper
are the more inclusive,
reporting longitude uncertainties as $5$--$10$ degrees of arc,
latitude uncertainties as $5$--$20$ degrees of arc,
and pole direction uncertainties as $10$--$20$ degrees of arc.
These ranges seem to be the best available information as
uncertainties also for
the sparse-data poles
reported in the subsequent four papers
that similarly do not include
individual error estimates
\citep{DURE16,DURE18a,DURE18b,DURE19},
for a total of thirteen of the objects
among those studied in the six aforementioned papers
that also are included in our observing program.
The RMS dispersion
of the individual latitude differences
between the program and sparse-data spin poles
is within the
reported range for the latitude uncertainties;
however,
the corresponding dispersion of $\pm 15$ degrees of arc
of the longitude differences is greater
than the range $5$--$10$ degrees of arc
estimated for the longitude uncertainties,
suggesting that the
reported range for the longitude coordinate errors of the sparse-data spin poles
is an underestimate.
For the remaining two observing program objects,
\citet{DURE20}
described a Monte-Carlo approach to estimate individual errors
for the pairs of pole locations that they reported.
50\% (two of four) of these longitudes
and
75\% (three of four) of these latitudes
are within
twice their estimated errors
from our corresponding results;
each of the
four poles is within
10 degrees of arc of our corresponding pole location.

\subsection{Spin vector sample}

The pole solutions for the observing program objects, 
together with the previously published results
summarized in Table~\ref{PREVDATA-TBL},
comprise a sample of 34 spin vectors in the Koronis family.
When sorted by decreasing brightness according to the
catalog absolute magnitudes \filter{H}
in Tables~\ref{LC-RESULTS-TBL} and \ref{PREVDATA-TBL}
the sample includes the brightest 32 family members,
complete to $H \approx 11.3$.
The two remaining program objects also had been
originally identified as part of the completeness sample,
but improvements to catalog absolute magnitudes during the years-long
observing program
have moved the objects relatively fainter in the brightness sequence
so that the sample at present includes 34 of the brightest 36 family members.
The stability of the bulk of the identifications of which objects
are the brightest members
supports confidence in the completeness of the sample.

\begin{table}
\caption{Previously published spin periods and obliquities
of Koronis family members
based on densely-sampled lightcurves.}
\label{PREVDATA-TBL}
\begin{tabular*}{\tblwidth}{@{}lrrrl@{}}
\toprule
Asteroid             &Catalog & Period & Spin    & Ref. \\
                     &\filter{H}\noteindex{a}&(h)&obliquity     &  \\
\midrule
 (158) Koronis       &  9.22      & 14.206     & 159$\degsym$ & b \\
 (167) Urda          &  9.23      & 13.061     & 163$\degsym$ & b \\
 (208) Lacrimosa     &  9.19      & 14.086     &  23$\degsym$ & c \\
 (243) Ida           &  9.93      &  4.634     & 156$\degsym$ & d \\
 (263) Dresda        & 10.23      & 16.814     &  16$\degsym$ & e \\
 (277) Elvira        &  9.90      & 29.691     & 167$\degsym$ & e \\
 (311) Claudia       & 10.00      &  7.532     &  50$\degsym$ & b \\
 (321) Florentina    & 10.12      &  2.871     & 154$\degsym$ & b \\
 (462) Eriphyla      &  9.38      &  8.659     &  51$\degsym$ & e \\
 (534) Nassovia      &  9.70      &  9.469     &  42$\degsym$ & e \\
 (720) Bohlinia      &  9.66      &  8.919     &  50$\degsym$ & b \\
 (832) Karin         & 11.29      & 18.352     &  42$\degsym$ & f \\
(1223) Neckar        & 10.59      &  7.821     &  47$\degsym$ & b \\
(1289) Kuta\"{\i}ssi & 10.65      &  3.624     & 165$\degsym$ & e \\
(1443) Ruppina       & 11.14      &  5.879     &  11$\degsym$ & g \\
\bottomrule
\end{tabular*}
\begin{flushleft}
\noteindex{a}
Catalog absolute magnitude
\filter{H} values as described for Table~\ref{LC-RESULTS-TBL}.
\newline
\noteindex{b}
\citet{SLIV03}
\newline
\noteindex{c}
\citet{VOKR21}
\newline
\noteindex{d}
\citet{BINZ93}
\newline
\noteindex{e}
\citet{SLIV09}
\newline
\noteindex{f}
\citet{SLIV12a}
\newline
\noteindex{g}
\citet{SLIV21b}
\end{flushleft}
\end{table}

The relative brightness sequence of
catalog absolute magnitudes is adopted here to represent
the sequence of the objects' relative sizes,
an approximation that is accurate enough for
the present discussion which does not depend on 
the exact sequence.

It is noted that 
the spin vector sample includes four asteroids that were
included in the observing program based on
identification as family members by \citet{MOTH05},
but are not identified in a subsequent family membership by \citet{NESV15b}:
(811) Nauheima, (1245) Calvinia, (1443) Ruppina, and (1848) Delvaux.
Mindful that determinations of
dynamical family membership involve in practice
an inexact process of distinguishing true members from
interlopers \citep{NESV15a},
these four objects are retained in the sample as likely real members
in the Koronis family based on their taxonomic classifications as S-type,
which spectrally distinguish them from the C-type local
background objects in the outer main belt.
Nauheima, Calvinia, and Delvaux are included as S-type in
the compilation by \citet{NEES10},
and
at the request of the first author of the present work
a reflectance spectrum of Ruppina was
obtained
using the SpeX instrument \citep{RAYN03}
for the SMASS Catalog of Asteroid
Spectra,\footnote{\tt http://smass.mit.edu/catalog.php}
which classifies\footnote{\tt http://smass.mit.edu/busdemeoclass.html}
as S-type in the Bus--DeMeo taxonomy \citep{DEME09}.
 
Koronis family member (832) Karin is bright enough to be in the sample,
but it is also the largest member of
the Karin cluster, a second-generation asteroid
family within the larger Koronis family \citep{CARR16}.
Because of its different spin evolution history,
Karin is excluded from the following discussion,
leaving 33 objects in the family analysis sample.

\subsection{Distributions of spin properties}

\begin{figure}
\centering
\includegraphics[scale=1.00]{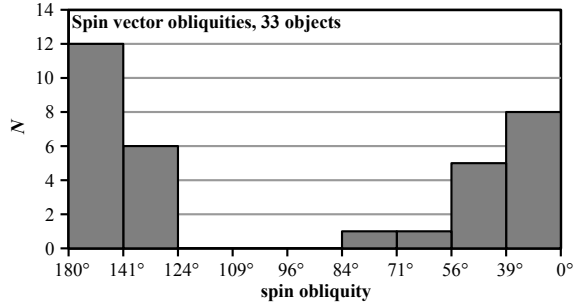}
\caption{Distribution of spin vector north pole obliquities
$\varepsilon$ in the analysis sample,
using bins of equal width in $\cos \varepsilon$ so that poles
isotropic on a sphere would give a uniform distribution.}
\label{OBLIQUITY-DISTRIB-FIG}
\end{figure}

A histogram of the spin vector obliquities derived for the sample is
shown in
Fig.~\ref{OBLIQUITY-DISTRIB-FIG}.
The small majority of retrograde spins is not statistically
significant;
using the binomial distribution the probability of randomly selecting
at least 18 objects with the same equally-likely spin direction from
of a sample of 33 objects $P(x\geq18$ or $x\leq15)$ is 73\%.
The overall bimodal shape of the distribution favoring low spin
obliquities persists from the previously-studied sample,
although two higher-obliquity prograde spins introduce some asymmetry.

\begin{figure}
\centering
\includegraphics[scale=1.00]{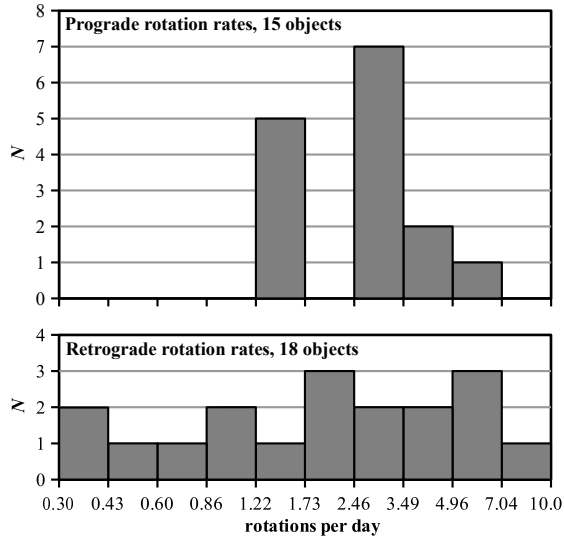}
\caption{Distributions of rotation rates in the
analysis sample
for the prograde- (upper graph) and retrograde-rotating
(lower graph) objects.}
\label{ROTRATE-DISTRIB-FIG}
\end{figure}

Histograms of the prograde and retrograde spin rates
for the objects in the analysis sample are shown in
Fig.~\ref{ROTRATE-DISTRIB-FIG}.
The spin rate distributions depend on the direction of spin because
the YORP evolution histories of
the prograde rotators can be subject to spin-orbit resonance trapping
that does not affect the retrograde rotators \citep{VOKR03}.
The expanded sample of
retrograde spins
(Fig.~\ref{ROTRATE-DISTRIB-FIG}, lower graph)
pushes the rotation rate lower bound to $\sim 0.4$~d$^{-1}$
and fills in the broad distribution of spin rates,
confirming that the
early impression of groupings in the retrograde rates
based on the incomplete previous sample
was an artifact.

In contrast,
the expanded prograde rotation rate distribution
(Fig.~\ref{ROTRATE-DISTRIB-FIG}, upper graph)
remains bimodal,
and also narrower overall than the retrograde distribution.
In particular,
no prograde rotation rates slower than 1.4~d$^{-1}$
are present,
whereas there are six retrograde objects having slower rotations
comprising one-third of the retrograde sample.

\begin{figure}
\centering
\includegraphics[scale=1.00]{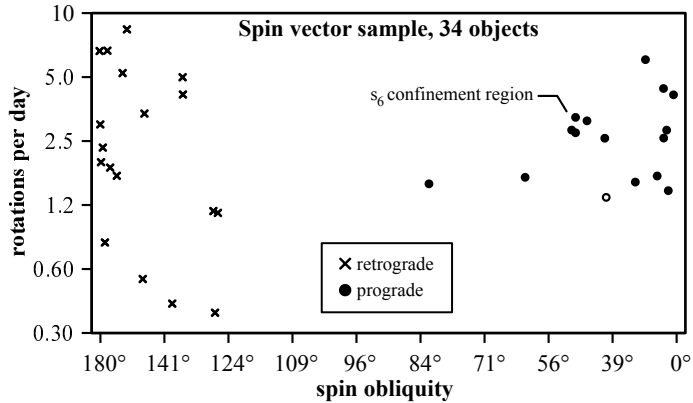}
\caption{Spin vector sample
as rotation rate vs. spin obliquity.
YORP torques modify objects' locations on this graph,
outward in both directions along the vertical axis
by rate spin-up or spin-down, and
outward in both directions along the horizontal axis
by reorienting spin vectors away from the orbits' planes.
The clustering of spin vectors near (45$\degsym$, 3.0 d$^{-1}$)
locates the confinement region for objects
trapped in the $s_6$ spin-orbit resonance \citep{VOKR03}.
The open circle symbol distinguishes
Karin cluster member (832) Karin
which is excluded from the analysis sample.}
\label{FREQ-VS-OBL-FIG}
\end{figure}

\citet{VOKR03} found for the previous study sample
that
the spin rates and obliquities distributions of the largest family members
are dominated by
modification
over 2--3 Gyr
by YORP thermal radiation torques.
YORP acts more quickly on smaller objects,
on more elongated objects,
and on
objects whose shapes have more ``windmill'' asymmetry;
thus to the extent that the objects' shapes (Fig.~\ref{SHAPES-FIG}) are
comparably irregular,
a more YORP-evolved
distribution among the smaller objects
can be expected.
The spin evolution modeling results
of \citet{VOKR03}
also suggested an expectation to find some
prograde objects
whose spin obliquity and spin rate significantly differ from the
clustered spin properties previously found by \citet{SLIV02} for
prograde Koronis members.
A scatter plot of rotation rates vs. spin obliquities of the
expanded sample of Koronis family spin vectors is shown
in Fig.~\ref{FREQ-VS-OBL-FIG},
on which is marked the confinement region for prograde spin
vectors trapped in the $s_6$ spin-orbit resonance.

Among the retrograde rotators,
the dominance of low spin obliquities
avoiding the orbit planes
suggests that the spins observed in the sample
are qualitatively consistent with 
YORP evolution modeling results,
with more-evolved spin rates
at the fast- and slow-rotating ends
of the distribution.
The prograde
distribution is more complicated because of the $s_6$
resonance already known,
and perhaps also
other possible resonances at longer periods and
smaller obliquities.

The three fastest-spinning prograde rotators in the expanded
sample---(811) Nauheima, (1443) Ruppina, and (2144) Marietta---are
smaller than all of the trapped $s_6$ objects
and have both shorter periods and smaller obliquities,
consistent with being further spun up by YORP.
Previously no prograde objects rotating faster than the trapped objects
had been identified,
although modeling had suggested that prograde rotators
whose shapes cause them 
predominately to spin up would be unlikely to be 
captured into the $s_6$
resonance,
and plausibly could evolve to fast prograde rotation instead
(D. Vokrouhlick\'{y}, personal communication).

Prograde rotators (1742) Schaifers and (2209) Tianjin
have spin periods similar to the trapped $s_6$ objects,
but are smaller in size.
Their respective spin obliquities of $17\degsym$ and $19\degsym$
are significantly less
than the $42\degsym$--$51\degsym$ obliquities for the known trapped objects,
and thus appear not to be trapped in $s_6$ themselves.

The five prograde-rotating analysis sample objects that
spin more slowly than the $s_6$-trapped objects
all share similar spin
periods of 14~h to 17~h,
but have spin obliquities
in a wide range from $16\degsym$ to $82\degsym$.
The largest of these objects is (208) Lacrimosa,
having size comparable to the largest of the trapped objects
but with a smaller spin obliquity of $23\degsym$.
Lacrimosa had been previously misclassified as retrograde spin
based on an
incorrect sidereal period not recognized as being an alias.
\citet{VOKR21}
determined its correct prograde spin vector solutions
and discuss how its
spin evolution differs from the other prograde objects of comparable size
which are trapped in the $s_6$ resonance,
concluding that it avoided capture in $s_6$ in its past
and also is not presently trapped in a longer-period resonance.

\citet{VOKR21}
also discuss possible evolution paths
including resonance capture and subsequent escape
that could have been experienced
by prograde-rotating objects smaller than Lacrimosa;
\citet{VOKR03} previously had found that prograde rotators that are
smaller or more irregular than the $s_6$ objects
that predominately spin down might already have evolved out
of the weakly unstable $s_6$ resonance,
and continued to spin down to other resonances with longer
characteristic rotation periods.
It could be instructive to explore such spin evolution possibilities
in the YORP model for
the two longer-period smaller prograde objects
in the expanded sample that have
smaller obliquities than the $s_6$-trapped objects,
(263) Dresda and (1029) La Plata
with spin obliquities of $16\degsym$ and $32\degsym$ respectively.

The remaining two smaller longer-period prograde rotators
(1100) Arnica and (1336) Zeelandia
have higher spin obliquities of
$62\degsym$ and $82\degsym$ respectively;
their spin axes are closer to their orbit planes than are those of any other
object of either spin direction in the analysis sample.

\subsection{Conclusions}

In this paper the lightcurve observations
and analyses for spins and convex shape models
of nineteen Koronis family members are reported,
which increase the sample of determined spin vectors in the Koronis family
to include 34 of the largest 36 family members,
and complete it
to $H \approx 11.3$ ($D\sim16$~km)
for the largest 32 members.
The pole locations reported here
agree on average with
poles for the same objects in the literature
that are based mainly
on lightcurves sparsely-sampled in time from sky surveys,
although the distribution of angular differences suggests that
the uncertainties of the sparse-data pole longitudes
are larger than the reported estimates.

Among the Koronis members in the completed size range,
the distributions of retrograde rotation rates and pole obliquities
appear to be qualitatively consistent with outcomes of modification by
thermal YORP torques.
The distribution of spin rates for the prograde rotators remains
narrower than that for the retrograde rotators;
in particular,
the absence of prograde rotators having periods longer than about 20 h
is real,
while among the retrograde rotators are several objects having longer
periods up to about 65 h.
None of the prograde objects newly added to the sample appear to be
trapped in an $s_6$ spin-orbit resonance
that is characteristic of most of the largest prograde objects
\citep{VOKR03};
these smaller objects either
could have been trapped previously and have already evolved out,
or have experienced spin evolution tracks that
did not include the resonance.

The results
reduce selection biases in the set of known spin vectors
for Koronis family members,
and will constrain future modeling of spin properties evolution.

\appendix
\section{Rotating a spin pole with respect to the
PGC pole}
\label{APDX-PGC}

The
3$\times$3 rotation matrix
${\bf R}(\theta,{\bf \hat{e}})$
that rotates
a vector by an angle $\theta$ about an axis direction
defined by the unit vector
${\bf \hat{e}} = (e_x, e_y, e_z)$
is given by,
for example,
\citet[p. 227]{COX86}
for which
positive rotation is counterclockwise
in a right-hand coordinate system.
Substituting for the components of
${\bf \hat{e}}$
in terms of
latitude $\phi$ and longitude $\lambda$,
\begin{align}
e_x & = \cos \phi \cos \lambda \nonumber \\
e_y & = \cos \phi \sin \lambda \\
e_z & = \sin \phi \nonumber 
\end{align}
yields
the elements of
the matrix
${\bf R}(\theta,\phi,\lambda)$
that rotates
a vector by an angle $\theta$ about an axis direction with
latitude $\phi$ and longitude $\lambda$:
\begin{equation}
{\bf R}(\theta,\phi,\lambda) = 
\begin{bmatrix}
R_{11} & R_{12} & R_{13} \\
R_{21} & R_{22} & R_{23} \\
R_{31} & R_{32} & R_{33}
\end{bmatrix}
\end{equation}
\begin{align*}
R_{11} & = 1-(1-\cos \theta ) (1-\cos^2 \lambda  \cos^2 \phi ) \\
R_{12} & = \sin \lambda  \cos \lambda  \cos^2 \phi  (1-\cos \theta )-\sin \phi  \sin \theta \\
R_{13} & = \cos \lambda  \sin \phi  \cos \phi  (1-\cos \theta )+\sin \lambda  \cos \phi  \sin \theta \\
R_{21} & = \sin \lambda  \cos \lambda  \cos^2 \phi  (1-\cos \theta )+\sin \phi  \sin \theta \\
R_{22} & = 1-(1-\cos \theta ) (1-\sin^2 \lambda  \cos^2 \phi )\\
R_{23} & = \sin \lambda  \sin \phi  \cos \phi  (1-\cos \theta )-\cos \lambda  \cos \phi  \sin \theta \\
R_{31} & = \cos \lambda  \sin \phi  \cos \phi  (1-\cos \theta )-\sin \lambda  \cos \phi  \sin \theta \\
R_{32} & = \sin \lambda  \sin \phi  \cos \phi  (1-\cos \theta )+\cos \lambda  \cos \phi  \sin \theta \\
R_{33} & = 1-\cos^2 \phi  (1-\cos \theta )
\end{align*}

The photometric great circle (PGC)
symmetry plane of observations
of an asteroid
is related to its orbit elements by
$\Omega_{\rm PGC} \approx \Omega$ and $i_{\rm PGC} \approx i(1+(1/2a))$
\citep{MAGN89}.
To rotate a spin pole with respect to the PGC pole,
first rotate the vector direction
${\bf \hat{p}}$
of the pole ecliptic coordinates into the PGC coordinate system by
\begin{equation}
{\bf \hat{q}} =
\left [ \begin{array}{c}
q_x \\
q_y \\
q_z \end{array} \right ]
= {\bf R}(-i_{\rm PGC},0,\Omega_{\rm PGC}){\bf \hat{p}}
\end{equation}
The rotated pole direction
${\bf \hat{r}}$
in ecliptic coordinates
that is $180\degsym$ away in PGC longitude
from
${\bf \hat{p}}$
is then
\begin{equation}
{\bf \hat{r}}
= {\bf R}(i_{\rm PGC},0,\Omega_{\rm PGC})\left [ \begin{array}{c}
-q_x \\
-q_y \\
q_z \end{array} \right ]
\end{equation}

\section*{Acknowledgments}

We thank
Edward Tedesco
and Claes-Ingvar Lagerkvist
for their invaluable help with
establishing the correct time tags of their
Koronis family lightcurve data from
more than 45 years ago,
David Vokrouhlick\'{y}
for helpful early discussions about YORP evolution scenarios,
and Jack Drummond
for calculating the moments of inertia results for the model shapes.

We thank
the corps of loyal observers who recorded data at Whitin Observatory:
Kathryn Neugent,
Amanda Zangari, 
Rebekah Dawson,
Kirsten Levan\-dow\-ski, 
Michaela Fendrock, 
Allison Youngblood, 
Carolyn Thay\-er, 
Kirsten Blancato, 
M.\ Claire Thoma, 
Anissa Benzaid, 
Katherine Lonergan, 
Molly Wass\-er, 
Kelsey Turbeville, 
Anne-Marie Hartt, 
Tim Smith, 
Alison Towner, 
Alejandra Escamilla Salda\~{n}a,
Ashley Iguina,
Leafia Sheraden Cox,
Christine Bachman, 
Gillian Beltz-Mohr\-mann,
Mariam Qazi, 
Rayna Rampalli,
Rebecca Stoll, 
Sormeh Yazdi, 
Emily Yax,
Rebecca Bernstein,
Naomi Gordon,
Karisa Zdanky,
Helen Ressler,
Chloe Shi,
Cassie Miller,
Floria Ngo,
and 
Elif Samanci.
At the Wallace Observatory
we thank Michael Person and Timothy Brothers for allocation
of telescope time and for observer instruction and support,
Timothy Brothers also for making some priority observations
in 2020
during the time that pandemic circumstances had closed the facility to visitors,
and the Wallace summer student observers in 2021:
Abigail Colclasure,
In\'{e}s Escobedo,
Aidan Henopp,
Rory Knight,
and
Andi Mitchell.

Participation of
coauthors
M. Hosek,
A. Sokol,
and M. Kurzner
was
supported by the National
Science Foundation under Grant Nos. AST-0647325 and AST-1005024.
Partial support for this work was provided by the U.S. Department of
Defense's Awards to Stimulate and Support Undergraduate Research
Education (ASSURE) program in collaboration with the National Science
Foundation's Research Experiences for Undergraduates program.
Coauthor S. Maynard,
observer R. Dawson,
and
observer K. Neugent
were supported by Sophomore Early Research Program grants
from the Wellesley College Office of the Dean of the College.
Student service observers at Whitin Observatory
were supported in part by grants from
the Massachusetts Space Grant Consortium.
The student observers
at Wallace Observatory
were supported by a grant from
MIT's
Undergraduate Research Opportunities Program. 
Use of the PROMPT telescopes for program observations
was made possible by support
from the Robert Martin Ayers Science Fund.
The observations made by coauthor A.\ Russell were funded by a
Theodore Dunham, Jr. Grant of the Fund for Astrophysical Research.
Coauthor F.\ Wilkin received funding from the Faculty Research Grant at Union
College.

The spectrum of
(1443) Ruppina
utilized for this publication was
obtained and made available by the MITHNEOS MIT-Hawaii Near-Earth
Object Spectroscopic Survey. The IRTF is operated by the
University of Hawaii under contract 80HQTR19D0030 with the
National Aeronautics and Space Administration. The MIT component
of this work is supported by NASA Grant 80NSSC18K0849.
The
taxonomic type result for the spectrum
was determined
using a Bus--DeMeo Taxonomy Classification Web tool
by Stephen M. Slivan, developed at MIT with the support of National
Science Foundation Grant 0506716 and NASA Grant NAG5-12355.

Opinions,
findings, and conclusions or recommendations expressed in this
material are those of the author(s) and do not necessarily reflect the
views of NASA or the National Science Foundation.

\section*{Data Availability}

Datasets related to this article can be found at
{\tt http://smass.mit.edu/slivan/lcdata.html}

\bibliographystyle{cas-model2-names}


\end{document}